\documentclass[12pt,english]{article}
\usepackage[T1]{fontenc}
\usepackage[latin9]{inputenc}
\usepackage[a4paper]{geometry}
\usepackage{array}
\usepackage{amsmath}
\usepackage{amssymb}
\usepackage{graphicx}

\makeatletter

\providecommand{\tabularnewline}{\\}

\numberwithin{equation}{section}

\usepackage{tikz}

\makeatother

\usepackage{babel}
\usepackage{listings}

\begin{document}

\title{The fuzzy space construction kit}

\author{Andreas Sykora\thanks{syko@gelbes-sofa.de}}

\date{October 3, 2016}
\maketitle
\begin{abstract}
Fuzzy spaces like the fuzzy sphere or the fuzzy torus have received
remarkable attention, since they appeared as objects in string theory.
Although there are many higher dimensional examples, the most known
and most studied fuzzy spaces are realized as matrix algebras defined
by three Hermitian matrices, which may be seen as fuzzy membranes
or fuzzy surfaces. We give a mapping between directed graphs and matrix
algebras defined by three Hermitian matrices and show that the matrix
algebras of known two-dimensional fuzzy spaces are associated with
unbranched graphs. By including branchings into the graphs we find
matrix algebras that represent fuzzy spaces associated with surfaces
having genus 2 and higher. \newpage{}
\end{abstract}
\tableofcontents{}

\newpage{}

\section{Introduction}

It is expected that space-time has a quantum structure at very short
distances. However, the concrete form of this quantum structure is
not known up to now. The different approaches to quantum gravity provide
different concepts. For example in loop quantum gravity, the states
are formed by so called spin networks, which are graphs having edges
labeled by representations of $SU(2)$ and nodes labeled with intertwining
operators that match the representations meeting at the respective
node. Another concept is non-commutative geometry, which tries to
generalize mathematical properties of ordinary manifolds to non-commutative
algebras. 

Some time ago, it was shown that also non-commutative structures emerges
from solutions of string theory based Yang-Mills matrix models such
as the IKKT and BFSS model \cite{Ishibashi:1997,Banks:1997}. Interestingly,
these structures are very similar to the structures developed in non-commutative
geometry. A compact manifold of dimension $d$ is replaced by a set
of $d$ finite dimensional matrices $X^{a}$, which may be seen as
the quantized coordinate functions of a non-commutative manifold. 

Up to now, only some examples of fuzzy spaces with very high symmetry
have been explicitly constructed and studied. All these spaces have
in common that they are associated with a classical manifold, which
is the commutative limit of a series of fuzzy spaces numbered by a
parameter. Recently, it was shown that also more general matrix algebras
may be associated with a classical manifold by finding a set of generalized
coherent states for the matrix algebra. In \cite{Ishiki:2015} this
set of coherent states was defined based on finding minimal energy
states of a quadratic Hamiltonian operator. In \cite{Berenstein:2012},
starting from specific effective Lagrangians derived from string theory,
a corresponding Hamiltonian similar to a Dirac operator was proposed,
and its zero modes were used for defining the classical manifold.
In \cite{Schneiderbauer:2016} it was shown that these two approaches
can be based on the same footing. In \cite{Badyn:2015} two-dimensional
surfaces emerging from the Hamiltonian defined in \cite{Ishiki:2015}
were studied in more detail. Interestingly, for the more symmetric
fuzzy spaces, which mainly are based on coadjoint orbits of Lie groups,
the set of coherent states are very similar to the coadjoint orbit
(seen as a manifold). These approaches makes it possible to study
less symmetric fuzzy spaces by investigating their set of coherent
states. Also fuzzy spaces without a classical limit may be studied.

In this work, we will give a quite general construction scheme for
matrix algebras based on three Hermitian matrices, which result in
surfaces of coherent states of arbitrary genus. Surfaces of higher
order genus were also considered in \cite{Arnlind:2009} without providing
an explicit representation for a genus higher than 1. 

In the following chapter 2, we mainly review the content of \cite{Berenstein:2012,Ishiki:2015,Badyn:2015,Schneiderbauer:2016},
which will be used in the following chapter 3. In general, the set
of coherent states can be defined based on a Laplace operator \cite{Ishiki:2015}
or a Dirac operator \cite{Berenstein:2012}. In the case of the Laplace
operator, a coherent state can be defined as state with minimal energy,
i.e. having a minimal eigenvalue for the Laplace operator. With the
Dirac operator, a coherent state can be defined based on the square
of the Dirac operator \cite{Schneiderbauer:2016}. The zero modes
of the Dirac operator, i.e. states with eigenvalue equal to zero,
represent a very interesting class of coherent states.

It is very important for the following that in the last case, it is
not necessary to determine the coherent states, when one is only interested
in the shape of the set of coherent states. Only the zero eigenvalues
of the Dirac operator have to be known. To this end, we will define
a functional, which is the determinant of the Dirac operator. Whenever,
the Dirac operator has a zero eigenvalue, this functional vanishes.
Since for finite-dimensional matrices the determinant is a polynomial
in the matrix entries, the set of coherent states or the surface of
zero modes is the manifold defined by the zero points of this polynomial.
The set of zero points of a polynomial easily can be visualized with
a computer program, which we mainly use to show the shapes of the
surfaces of zero modes.

In chapter 3, we then define a mapping from directed graphs with $N$
nodes to three Hermitian $N\times N$-matrices. For many known fuzzy
spaces, such as the fuzzy sphere, the fuzzy torus and the fuzzy plane,
we construct the corresponding graphs. We will see that all these
fuzzy spaces are based on unbranched graphs. By generalizing to branched
graphs, we then realize fuzzy spaces of higher genus (or at least
with zero modes surfaces of higher genus).

\section{Quasi-coherent states}

In this chapter we mainly review the approach of investigating a fuzzy
space based on coherent states as put forward in \cite{Berenstein:2012,Ishiki:2015,Badyn:2015,Schneiderbauer:2016}. 

A fuzzy space is defined via a set of $d$ Hermitian $N\times N$-matrices
$X^{a},$ $a=1,\dots,d$, which can be interpreted as the quantized
embedding functions $x^{a}$ of a classical manifold embedded in $\mathbb{R}^{d}$.
There are numerous examples for such spaces, such as the fuzzy sphere
or the fuzzy torus. It is also possible that the matrices are infinite
dimensional, which, for example, is the case for the fuzzy plane.
In general, such matrices can be seen as Hermitian operators of an
possibly infinite dimensional Hilbert space. 

The algebra generated by the matrices $X^{a}$ can be interpreted
as the non-commutative version of the function algebra of the fuzzy
space. There is a correspondence of states $\left|\psi\right\rangle $
of the Hilbert space the matrix algebra acts on, and the elements
of the matrix algebra. For every normalized state $\left|\psi\right\rangle $
there is an projector $\left|\psi\right\rangle \left\langle \psi\right|$
in the matrix algebra. On the other hand for every projector $p$
with $p^{2}=p$ in the matrix algebra, there is a state $\left|\psi\right\rangle $
with $p_{\psi}=\left|\psi\right\rangle \left\langle \psi\right|$.
Coherent states can be interpreted as generalization of $\delta-$functions,
but we do not explore this direction further.

\subsection{Coherent states}

According to Perelomov \cite{Perelomov:1972} coherent states can
be defined on representations of Lie groups. On more general fuzzy
spaces, without the action of a group, the notion of coherent states
can be generalized, for example to ``quasi-coherent states\textquotedbl{}
(see\cite{Schneiderbauer:2016}) . A quasi-coherent state is defined
as a state with minimal dispersion and maximal localization. 

To be specific, for every state $\left|\psi\right\rangle $ we can
calculate the expectation value of the matrices $X^{a}$
\begin{equation}
\left\langle X^{a}\right\rangle _{\psi}=\left\langle \psi\right|X^{a}\left|\psi\right\rangle =\mathrm{tr}\left(X^{a}p_{\psi}\right)
\end{equation}
and based on the standard deviation 
\begin{equation}
\left(\Delta_{\psi}(X^{a})\right)^{2}=\left\langle \psi\right|\left(X^{a}\right)^{2}\left|\psi\right\rangle -\left\langle \psi\right|X^{a}\left|\psi\right\rangle ^{2}
\end{equation}
the dispersion $\delta_{\psi}$ of a state may be defined
\begin{equation}
\delta_{\psi}=\sum_{a}\left(\Delta_{\psi}(X^{a})\right)^{2}
\end{equation}

In \cite{Schneiderbauer:2016} it is proposed to restrict to states,
which have a low dispersion and are localized at a specific point
$x$, meaning that the expectation values $\left\langle \psi\right|X^{a}-x^{a}\left|\psi\right\rangle $
are low in some sense. For example, both dispersion and localization
can be optimized simultaneously by minimizing the function 
\begin{equation}
E_{\Delta}(x)=\left|x^{a}-\left\langle X^{a}\right\rangle _{\psi}\right|^{2}+\delta_{\psi}\label{eq:laplace string energy}
\end{equation}
which can be interpreted as the energy of a string attached to the
fuzzy space and extending to the point $x$. This energy can be derived
by defining an extended coordinate function on a background space

\begin{equation}
\mathfrak{X}^{a}=\left(\begin{array}{cc}
X^{a}\\
 & x^{a}
\end{array}\right)
\end{equation}
which can be interpreted as the fuzzy space (or non-commutative brane)
together with a point probe at the point $x^{a}$. The off-diagonal
entries 

\begin{equation}
\Psi=\left(\begin{array}{cc}
 & \left|\psi\right\rangle \\
\left\langle \psi\right|
\end{array}\right)\label{eq:strings_pointprobe}
\end{equation}
on the background space can be interpreted as strings interconnecting
the point probe with the fuzzy space. On the background space, an
extended Laplace operator $\Delta_{\mathfrak{X}}$ can be defined,
which when applied to the ``strings'' results in 
\begin{equation}
\Delta_{\mathfrak{X}}\Psi=\delta_{ab}[\mathfrak{X}^{a},[\mathfrak{X}^{a},\Psi]]=\left(\begin{array}{cc}
 & \Delta_{x}\left|\psi\right\rangle \\
\left\langle \psi\right|\Delta_{x}
\end{array}\right)
\end{equation}
where $\Delta_{x}$ is a localized Laplace operator

\begin{equation}
\Delta_{x}=\sum_{a}(X^{a}-x^{a})^{2}
\end{equation}
The localized Laplace operator has the property that $\left\langle \psi\right|\Delta_{x}\left|\psi\right\rangle =E_{\Delta}(x)$
or $\Delta_{x}\left|\psi\right\rangle =E_{\Delta}(x)\left|\psi\right\rangle $
with $E_{\Delta}(x)$ as defined in (\ref{eq:laplace string energy}).
Thus, the quasi-coherent states are ground states of the localized
Laplace operator and in principle, the quasi-coherent states can be
determined by solving an eigenvalue problem.

As an alternative, quasi-coherent spinor states can be considered
\cite{Schneiderbauer:2016,Berenstein:2012}. On the background space
with the point probe, also a Dirac operator $D_{\mathfrak{X}}$ can
be defined, which when acting on spinorial, off-diagonal ``strings''
results in a localized Dirac operator 
\[
D_{\mathfrak{X}}=\gamma_{a}[\mathfrak{X}^{a},\Psi]=\left(\begin{array}{cc}
 & D_{x}\left|\psi\right\rangle \\
\left\langle \psi\right|D_{x}
\end{array}\right)
\]
Here, the $\gamma_{a}$ are the $[\frac{d}{2}]$-dimensional generators
of the Clifford algebra associated with $\mathbb{R}^{d}$, $\Psi$
is a spinor valued matrix on the background and $\left|\psi\right\rangle $
is a spinor valued vector. The localized Dirac operator is 
\begin{equation}
D{}_{x}=\sum_{a}\gamma_{a}(X^{a}-x^{a})\label{eq:localized Dirac}
\end{equation}
 The quasi-coherent spinor states can then be defined as the ground
states of

\begin{equation}
D_{x}^{2}\left|\psi\right\rangle =E_{D}(x)\left|\psi\right\rangle 
\end{equation}
Note that the string energy $E_{D}(x)$ defined by the square of the
localized Dirac operator $D{}_{x}$ differs from the string energy
$E_{\Delta}(x)$ of the localized Laplace operator $\Delta_{x}$ (see
\cite{Schneiderbauer:2016}).

When one has defined quasi-coherent states, every point $x$ of $\mathbb{R}^{d}$
can be assigned to the minimal string energy $E(x)$ of the localized
Laplace operator $\Delta_{x}$ or the localized Dirac operator $D_{x}$
at that point $x$. 
\begin{equation}
E(x)=\min\left(\mathrm{spec}(\Delta_{x}\,\mathrm{or}\,D_{x})\right)
\end{equation}
The function $E$ is differentiable nearly everywhere, expect at points,
where minimal eigenvalues cross. At points, where the minimal string
energy $E$ is much higher than an overall minimal string energy,
we expect that the end of the string located at $x^{a}$ is far away
from the fuzzy space defined by the matrices $X^{a}$. On the other
hand, when the minimal string energy $E$ is small, we expect that
the string is short and that we are near the fuzzy space. Due to (\ref{eq:laplace string energy}),
we see that when the end of the string is moved away from the fuzzy
space, the energy should rise at least quadratically. A computer program
for determining point sets with ``small'' string energies, which
visualizes this behavior, is described in \cite{Schneiderbauer:2016}.

\subsection{Zero modes manifolds of the localized Dirac operator}

In \cite{Berenstein:2012,Badyn:2015,Schneiderbauer:2016} the observation
was made that nearly all known fuzzy spaces not only have coherent
states defined by (\ref{eq:laplace string energy}) but also - which
at a first glance seems to be more restrictive - defined by the zero
modes of (\ref{eq:localized Dirac}). In particular, when the localized
Dirac operator $D_{x}$ (\ref{eq:localized Dirac}) has a zero mode
at the point $x$, i.e. the corresponding quasi-coherent state is
an eigenvector with eigenvalue $0$: $D_{x}\left|\psi\right\rangle =0$,
then $D_{x}^{2}\left|\psi\right\rangle =0$ and the (never negative)
string energy defined by the positive definite Hermitian operator
$D_{x}^{2}$ is minimal at this point. 

When the matrices defining the fuzzy space are finite dimensional,
the above statement is equivalent to
\begin{equation}
\det D_{x}=\det\left(\gamma_{a}(X^{a}-x^{a})\right)=0\label{eq:det_dirac}
\end{equation}
as was already noted in \cite{Badyn:2015}. The left hand side of
this equation defines a multivariate polynomial of the $d$ variables
$x^{a}$ of order $q=2[\frac{d}{2}]N$, where $[\cdot]$ takes the
integer of its argument. Considering the term $-\gamma_{a}x^{a}$
inside the determinant (\ref{eq:det_dirac}), since the $\gamma_{a}$
contain an entry $\pm1$ or $\pm i$ in every row and an even number
of minus signs, the determinant comprises $\pm\sum_{a}(x_{a})^{q}$
as term of highest order. Thus, as a function, the polynomial goes
to $\pm$infinity for $\left|x\right|$ going to infinity. If there
is at least one point at which the polynomial is smaller or bigger
than $0$, opposite to the sign of the highest term, then there has
to be a submanifold that surrounds this point, on which the polynomial
vanishes. In the following, we will call this manifold ``zero modes
manifold'' or ``zero modes surface'' in the case of a two-dimensional
manifold.

The determinant functional (\ref{eq:det_dirac}) has similar properties
as the index function defined in \cite{Berenstein:2012}:
\begin{itemize}
\item It is invariant under unitary transformations
\begin{equation}
X^{a}\longrightarrow UX^{a}U^{\dagger}\label{eq:unitary_trafo}
\end{equation}

\item It is covariant under translations, rotations and scaling 
\begin{equation}
X^{a}\longrightarrow\alpha R_{b}^{a}X^{b}+c^{a}
\end{equation}

\item If the matrices are block diagonal, i.e. the fuzzy spaces is defined
by the direct sum of two smaller fuzzy spaces 
\begin{equation}
X^{a}=\left(\begin{array}{cc}
X_{1}^{a}\\
 & X_{2}^{a}
\end{array}\right)
\end{equation}
the determinant splits 
\begin{equation}
\det D_{x}=\det\left(\gamma_{a}(X_{1}^{a}-x^{a})\right)\det\left(\gamma_{a}(X_{2}^{a}-x^{a})\right)=0
\end{equation}
which is equivalent to 
\begin{equation}
\det\left(\gamma_{a}(X_{1}^{a}-x^{a})\right)=0\;\mathrm{or}\;\det\left(\gamma_{a}(X_{2}^{a}-x^{a})\right)=0
\end{equation}
In this case, the overall zero modes manifold is the combine of the
zero modes manifolds of the fuzzy spaces defined by $X_{1}^{a}$ and
$X_{2}^{a}$.
\end{itemize}
The invariance with respect to unitary transformations has interesting
implications. The transformation (\ref{eq:unitary_trafo}) can be
seen as non-commutative symplectic coordinate transformation, since
for $Y^{a}$$=UX^{a}U^{\dagger}$ a function on the fuzzy space transforms
like $U\hat{f}(X^{a})U^{\dagger}=\hat{f}(Y^{a}).$ Furthermore, since
explicitly $0=\det\left(\gamma_{a}(X^{a}-x^{a})\right)=\det\left(\gamma_{a}(Y^{a}-x^{a})\right)$,
the matrices $X^{a}$ and $Y^{a}$ have the same zero modes manifold.

To get used to the determinant functional (\ref{eq:det_dirac}) ,
we study it in low space dimensions and low matrix dimensions:
\begin{itemize}
\item In one matrix dimension, i.e. $N=1$, equation (\ref{eq:det_dirac})
reduces to $X^{a}=x^{a}$, i.e. the one-dimensional matrices $X^{a}$
define a point in $\mathbb{R}^{d}$. \\
There is an interesting consequence, when a fuzzy space is defined
by $d$ commuting $N$-dimensional matrices $X^{a}$. In this case,
the matrices $X^{a}$ are simultaneously diagonalizable, the determinant
functional (\ref{eq:det_dirac}) is invariant with respect to the
transformation necessary for diagonalizing the matrices, and due to
the splitting property, the matrices $X^{a}$ define $N$ points in
$\mathbb{R}^{d}$. 
\item In one space dimension, $d=1,$ there is only one matrix $X$, the
matrix can be diagonalized and the eigenvalues define $N$ points
on the line $\mathbb{R}$.
\item In two space dimensions $d=2$, with two matrices $X$ and $Y$, equation
(\ref{eq:det_dirac}) reduces to
\begin{equation}
\det\left(\begin{array}{cc}
0 & (X-x)-i(Y-y)\\
(X-x)+i(Y-y) & 0
\end{array}\right)=0\label{eq:2D-det_dirac}
\end{equation}
Thus with $V=X+iY$ and $v=x+iy$ we arrive at 
\begin{equation}
\det(V-v)=0
\end{equation}
The zero modes manifold of this fuzzy space is defined by the complex
eigenvalues of the matrix $V$.
\end{itemize}
In the last case with $d=2$, there is an example, the fuzzy disc
\cite{Lizzi:2003}, which can be described by $V=\sum_{n=0}^{N-1}\sqrt{n+1}\left|n\right\rangle \left\langle n+1\right|$,
i.e. an upper triangular matrix. For every upper triangular matrix
$\det(V-v)=(-v)^{N}$, i.e. the eigenvalues are all zero. The zero
modes manifold is only the origin. However, equation (\ref{eq:2D-det_dirac})
is equal to $(x^{2}+y^{2})^{N}$. I.e. for large $N$, inside the
circle with radius $1$, it is almost zero and outside of the circle
it is very large. In this context, this justifies to speak of a fuzzy
disc. However, the fuzzy disc does not have a surface as zero modes
manifold.

In the remaining, we will consider three space dimensions $d=3$,
with three matrices $X,$ $Y$ and $Z$. Equation (\ref{eq:det_dirac})
becomes
\begin{equation}
\det\left(\begin{array}{cc}
Z-z & (X-x)-i(Y-y)\\
(X-x)+i(Y-y) & -(Z-z)
\end{array}\right)=\det\left(\begin{array}{cc}
Z-z & V^{\dagger}-\bar{v}\\
V-v & -(Z-z)
\end{array}\right)=0\label{eq:3D-det_dirac}
\end{equation}
where we have set $V=X+iY$and $v=x+iy$.

\section{Graphs and matrices}

In this chapter, we give a mapping between directed graphs and three
matrices $X,$ $Y$ and $Z$, which, at least in the case of sparse
matrices, will result in a fuzzy space that (substantially) has the
same topology as the graph. We will use the determinant functional
(\ref{eq:det_dirac}) to determine the zero modes surface of specific
graphs and will show, how to define graphs that are mapped to fuzzy
spaces with a desired topology of the zero modes surface.

\subsection{The basic building block\label{sub:The-basic-building}}

The basic building block is a graph with two nodes and one interconnecting
edge. We label the two nodes with two points in $\mathbb{R}^{3}$,
i.e. $(x_{1},y_{1},z_{1})$ and $(x_{2},y_{2},z_{2})$. The edge is
labeled with two complex numbers $s_{x}$ and $s_{y}$. This graph
is mapped to the matrices 

\begin{equation}
X=\left(\begin{array}{cc}
x_{1} & s_{x}\\
\overline{s_{x}} & x_{2}
\end{array}\right),\ Y=\left(\begin{array}{cc}
y_{1} & -is_{y}\\
i\overline{s_{y}} & y_{2}
\end{array}\right),\ Z=\left(\begin{array}{cc}
z_{1}\\
 & z_{2}
\end{array}\right)
\end{equation}

When the numbers $s_{x}$ and $s_{y}$ are real, we will see that
a very helpful visualization is that the two points are interconnected
with a closed string that emerges from the first point, blows up to
an ellipse with radii $s_{x}$ and $s_{y}$ in the $x$- respective
$y$-direction and shrinks to the second point. A further analogy
comes from the comparison with (\ref{eq:strings_pointprobe}): The
numbers $s_{x}$ and $s_{y}$ are the extreme case of open strings
interconnecting the two point-probes $(x_{1},y_{1},z_{1})$ and $(x_{2},y_{2},z_{2})$.

It is also possible to map the Hermitian matrices $X$ and $Y$ to
one complex matrix $V$ via $V=X+iY$. The coordinates for $x$ and
$y$ can be provided with two complex numbers $v_{1}=x_{1}+iy_{1}$
and $v_{2}=x_{1}+iy_{2}$ and the ``string radii'' with two complex
numbers $s_{12}=s_{x}+s_{y}$ and $s_{21}=\overline{s_{x}}-\overline{s_{y}}$.
The matrices defined by the graph of the basic building block are
then

\begin{equation}
V=\left(\begin{array}{cc}
v_{1} & s_{12}\\
s_{21} & v_{2}
\end{array}\right),\,Z=\left(\begin{array}{cc}
z_{1}\\
 & z_{2}
\end{array}\right)
\end{equation}

Fig. 1 shows the graph of the basic building block on the left side.
On the right side a zero modes surface with $\vec{x}_{1}=(0,0,-1/2)$,
$\vec{x}_{2}=(0,0,1/2)$, $s_{x}=\alpha/2$ and $s_{y}=\alpha/2$
for $\alpha=1/2$ is shown, which has been plotted with a simple computer
program that is described in appendix \ref{sec:SageMath-program}.
Basically, the computer program approximately determines a surface
in $\mathbb{R}^{3}$, where (\ref{eq:det_dirac}) is fulfilled.

\begin{figure}

\begin{tabular}{>{\centering}p{7cm}>{\centering}m{7cm}}
\begin{tikzpicture}
	\tikzstyle{N}=[draw,circle,fill=white,minimum size=4pt,inner sep=0pt]

	\node[N] (n1) at (0pt,0pt)[label=right:$\left( x_1\ y_1\ z_1 \right)$] {}; 

	\node[N] (n2) at (0pt,20pt) [label=right:$\left( x_2\ y_2\ z_2 \right)$] {}
		edge[<-] node[anchor=east] {$\left( s_x\ s_y \right)$} (n1);

\end{tikzpicture} & \includegraphics[scale=0.4]{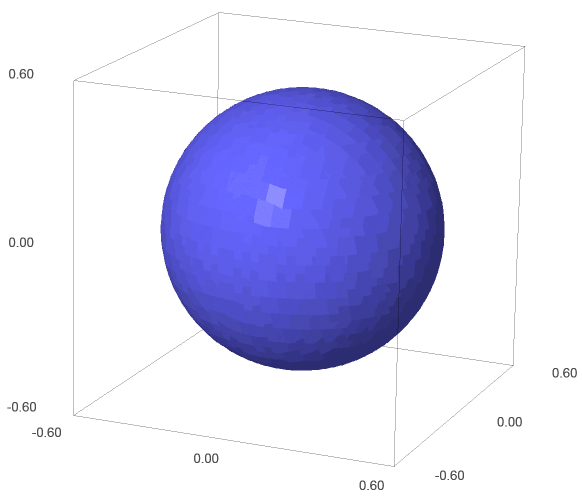}\tabularnewline
\end{tabular}\protect\caption{Basic building block}

\end{figure}
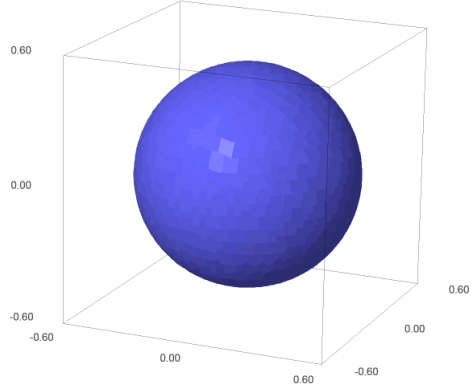

For the example on the right of Fig. 1, the coordinates and string
radii are mapped to

\begin{equation}
X=\frac{1}{2}\left(\begin{array}{cc}
0 & 1\\
1 & 0
\end{array}\right)=\frac{1}{2}\sigma_{1},\:Y=\frac{i}{2}\left(\begin{array}{cc}
0 & -1\\
1 & 0
\end{array}\right)=\frac{1}{2}\sigma_{2},\:Z=\alpha\left(\begin{array}{cc}
1\\
 & -1
\end{array}\right)=\alpha\sigma_{3}
\end{equation}
The factor $\frac{1}{2}$ has been chosen such that $V$ is simply
$\left(\begin{array}{cc}
0 & 1\\
0 & 0
\end{array}\right)$.

For this example with arbitrary $\alpha$, it is easy to analytically
determine the determinant functional, which results in

\begin{equation}
p=\det\left(\gamma_{a}(X^{a}-x^{a})\right)=r^{4}+z^{4}+2r^{2}(z^{2}+\alpha^{2})+z^{2}(1-2\alpha^{2})+\alpha^{4}-\alpha^{2}
\end{equation}
with $r^{2}=x^{2}+y^{2}$. 

The zero modes surface has a rotational symmetry with respect to $x$
and $y$. For $r=0$ we can determine the coordinates for $z$ via
\begin{equation}
p(r=0)=\left(z^{2}-\alpha^{2}\right)\left(z^{2}-\alpha^{2}+1\right)
\end{equation}
The $\alpha$-dependent family of zero modes surfaces shows many interesting
features, which arise also for the more complicated fuzzy spaces constructed
below.
\begin{itemize}
\item For $0<\alpha<\frac{1}{2}$ the zero modes surface has the shape of
a quenched sphere or disc.
\item For $\alpha=\frac{1}{2}$, the zero modes surface is a (real and not
only topological) sphere with radius $\frac{1}{2}$ and the polynomial
$p$ decomposes into
\begin{equation}
p=\left(r^{2}+z^{2}-\frac{1}{4}\right)\left(r^{2}+z^{2}+\frac{3}{4}\right)
\end{equation}
(In this case, the matrices define the smallest fuzzy sphere.)
\item For $\frac{1}{2}<\alpha<1$ the zero modes surface gets more and more
narrower in the middle, like a barbell.
\item For $\alpha=1$, the zero modes surface has a shape like a 2p-orbital
along the $z$-axis or equivalently like a 8-curve rotated about the
$z$-axis.
\item For $\alpha>1$, the zero modes surface is composed of two disconnected
topological spheres, which with increasing $\alpha$ shrink and are
more and more remote from each other.
\end{itemize}

\subsection{The general mapping}

In general, we have a graph with $N$ nodes, which are labeled with
$N$ points $(x_{i},y_{i},z_{i})$. The nodes are numbered, however
the following mapping to matrices is covariant under permutations
of the numbering of the nodes in the sense that a permutation of nodes
results in a permutation of matrix row and line numbers. These permutations
are a subgroup of the matrix transformation group under which $\det D_{x}$
is invariant. More general, a permutation of nodes corresponds to
a permutation of a basis of the Hilbert space, the matrices act on.

As second ingredient, the graph has directed edges, which interconnect
the nodes (only at least one edge interconnects two nodes). The edges
are labeled with two complex numbers $(s_{xij},s_{yij})$ , where
$i$ and $j$ are the number of the nodes, which are interconnected
by the edge.

A graph defined in this way is mapped to three quadratic matrices
$X,Y,Z$ of dimension $N$, where we demand that the $2\times2$ sub-matrices
of $X$, $Y$, and $Z$ corresponding to one edge have the form of
the basic building block described above. In particular:
\begin{itemize}
\item The diagonal entries of the matrices are filled with the coordinates
of the points, i.e. 
\begin{equation}
X_{ii}=x_{i},\ Y_{ii}=y_{i},\:Z_{ii}=z_{i}
\end{equation}

\item When the graph has an edge from node $i$ to node $j$, the off-diagonal
elements of $X$ and $Y$are filled according to 
\begin{equation}
X_{ij}=s_{xij},\,X_{ij}=\overline{s_{xij}},\quad Y_{ij}=-is_{xij},\,Y_{ji}=i\overline{s_{xij}}
\end{equation}

\item The other entries of the matrices are zero.
\end{itemize}
The matrices $X$, $Y$ and $Z$ are Hermitian. Note that two graphs
with the same nodes and edges but different edge directions are not
equal and will not result in equal zero modes surfaces. Changing the
direction of an edge corresponds to transposing the corresponding
basic building block. 

When using the matrix $V=X+iY$, the mapping is
\begin{equation}
V_{ii}=x_{i}+iy_{i}=v_{i},\:Z_{ii}=z_{i}
\end{equation}
\begin{equation}
V_{ij}=s_{ij}=s_{xij}+s_{yij},\;V_{ji}=s_{ji}=\overline{s_{xij}}-\overline{s_{yij}}
\end{equation}

Here, changing the direction of an edge from $i$ to $j$ corresponds
to exchanging $s_{ij}$ with $s_{ji}$.

For three arbitrary Hermitian matrices it is also possible to provide
an inverse mapping that results in a graph with a set of matrices
that are equivalent to these matrices up to a unitary transformation.
In particular, the three arbitrary matrices and the matrices of the
graph have the same zero modes surface:

Since the three arbitrary matrices are Hermitian, one matrix, say
$Z$, can be diagonalized with a unitary transformation. The other
two matrices $X$ and $Y$ can be transformed with this unitary transformation.
From the transformed matrices $X$ and $Y$ and the diagonalized matrix
$Z$, the diagonal entries form the points of the graph. For every
non-zero diagonal entry of $X$ and $Y$ an edge has to be added to
the graph. The labels $s_{x}$ and $s_{y}$ of the edge can be read
off from the corresponding matrix entries. Due to the choice of the
matrix, which is diagonalized, there are in principle three different
graphs associated with three Hermitian matrices.

Interestingly, after diagonalizing $Z$, what remains from the symmetry
of unitary transformations are permutations of the eigenvectors (which
are basis vectors for the space, on which the matrices act) and phase
transformations of the eigenvectors. The permutation of eigenvectors
corresponds to a permutation of a numbering of the nodes of the graph,
i.e. the mapping is well-defined for graphs with unnumbered nodes.

The phase transformations of the eigenvectors can be encoded with
diagonal unitary matrices $U$, i.e. matrices that only have diagonal
entries $U_{ii}=e^{i\varphi_{i}}$ with real parameters $\varphi_{i}$.
When one transforms the matrices $X$, $Y$ and $Z$ with such a transformation,
one sees that the diagonal entries, i.e. the points associated to
the nodes, are invariant under these transformations and that the
entries for an edge from node $i$ to node $j$ acquire a phase \-$e^{i(\varphi_{i}-\varphi_{j})}$.
Such a unitary transformation therefore can be seen at a first glance
as a local $U(1)$ lattice gauge transformation defined on the nodes
and acting on the ends of the edges. However, since all unitary transformations
correspond to symplectomorphism, also these transformations correspond
to coordinate transformations of the zero modes manifold.

In the following, we will illustrate the mapping in more and more
complex examples. We will show that, when the entries of the matrices
are all substantially of the same magnitude, the graphs define fuzzy
spaces with zero modes surfaces that have an analogous topology as
the graph. In such a way, we are able to define fuzzy spaces with
zero modes surfaces of arbitrary genus.

\subsection{Fuzzy cylinder}

The algebra relations 
\begin{equation}
VV^{\dagger}=V^{\dagger}V=r_{0}^{2}1,\quad[Z,V]=V
\end{equation}
define an infinite fuzzy cylinder (for example, see \cite{Steinacker:2011})
with radius $r_{0}$. A matrix representation of this algebra is 
\begin{equation}
V=\sum_{n\in\mathbb{Z}}r_{0}\left|n\right\rangle \left\langle n+1\right|,\quad Z=\sum_{n\in\mathbb{Z}}n\left|n\right\rangle \left\langle n\right|
\end{equation}
In Fig. 2, the corresponding graph is shown on the left. In the graph
we have indicated the $z_{n}$ of the points to the left of the nodes.
The $x_{n}$ and $y_{n}$ are all zero. In the below, we will omit
zero coordinates and will try to align the nodes of the graphs according
to their $z$-coordinate. Furthermore, we have introduced a notation
with only one number $s$ labeling an edge. This means that for $i<j$
the corresponding $s_{ij}=s$ and $s_{ji}=0$.

\begin{figure}
\begin{tabular}{>{\centering}m{4.5cm}>{\centering}m{4.5cm}>{\centering}m{5cm}}
\begin{tikzpicture}
	\tikzstyle{N}=[draw,circle,fill=white,minimum size=4pt,inner sep=0pt]

	\node at (40pt,-60pt) {$\vdots$}; 
	\node at (40pt,-40pt) {$-2$}; 
	\node at (40pt,-20pt) {$-1$}; 
	\node at (40pt,0pt)   {$0$}; 
	\node at (40pt,20pt)  {$1$};
	\node at (40pt,40pt)  {$2$}; 
	\node at (40pt,60pt) {$\vdots$}; 

	\node at (60pt,-60pt) {$\vdots$}; 
	\node[N] (n1) at (60pt,-40pt) {}; 
	\node[N] (n2) at (60pt,-20pt) {} edge[<-] (n1);
	\node[N] (n3) at (60pt,0pt)  {} edge[<-] (n2);
	\node[N] (n4) at (60pt,20pt) {} edge[<-] (n3);
	\node[N] (n5) at (60pt,40pt) {} edge[<-] (n4);
	\node at (60pt,60pt) {$\vdots$}; 

	\node at (80pt,-30pt) {$s=r_0$}; 
	\node at (80pt,-10pt) {$s=r_0$}; 
	\node at (80pt,10pt)  {$s=r_0$}; 
	\node at (80pt,30pt)  {$s=r_0$}; 

\end{tikzpicture} & \begin{tikzpicture}
	\tikzstyle{N}=[draw,circle,fill=white,minimum size=4pt,inner sep=0pt]

	\node at (40pt,-40pt) {$0$}; 
	\node at (40pt,-20pt) {$1$}; 
	\node at (40pt,0pt)   {$2$}; 
	\node at (40pt,20pt)  {$3$};
	\node at (40pt,40pt)  {$4$}; 

	\node[N] (n1) at (60pt,-40pt) {}; 
	\node[N] (n2) at (60pt,-20pt) {} edge[<-] (n1);
	\node[N] (n3) at (60pt,0pt)  {} edge[<-] (n2);
	\node[N] (n4) at (60pt,20pt) {} edge[<-] (n3);
	\node[N] (n5) at (60pt,40pt) {} edge[<-] (n4);

	\node at (80pt,-30pt) {$s=r_0$}; 
	\node at (80pt,-10pt) {$s=r_0$}; 
	\node at (80pt,10pt)  {$s=r_0$}; 
	\node at (80pt,30pt)  {$s=r_0$}; 

\end{tikzpicture} & \includegraphics[scale=0.5]{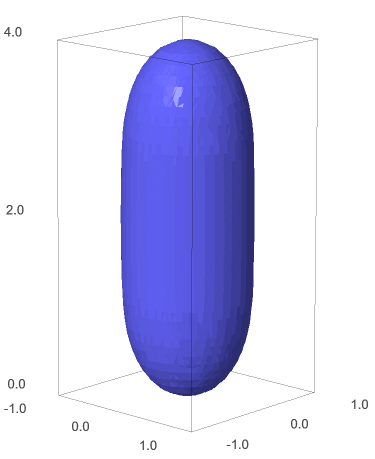}\tabularnewline
\end{tabular}\protect\caption{Fuzzy cylinder}

\end{figure}
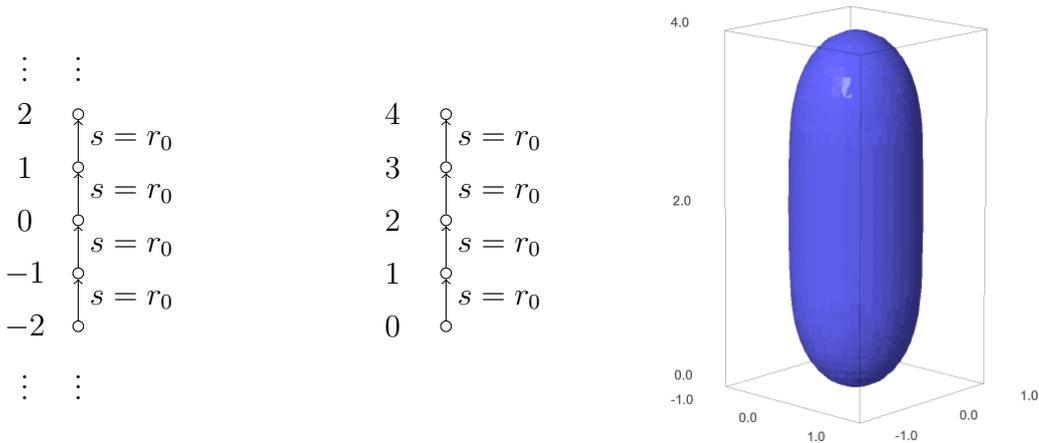

The graph in the middle corresponds to a finite fuzzy cylinder of
length 5, which in general can be defined by projecting out a finite
part of the infinite cylinder with a projector $P=\sum_{n=0}^{N-1}\left|n\right\rangle \left\langle n\right|$.
The matrices $Z_{P}=PZP$ and $V_{P}=PVP$ for the finite fuzzy cylinder
can be represented by
\begin{equation}
V_{P}=r_{0}\sum_{n=0}^{N-2}\left|n\right\rangle \left\langle n+1\right|,\;Z_{P}=\sum_{n=0}^{N-1}n\left|n\right\rangle \left\langle n\right|
\end{equation}

To the right of Fig. 2 we have included a picture of the zero modes
surface of the middle graph with $s=1$. Interestingly, fuzzy spaces
seem not to distinguish between open and closed ``manifolds''. Although
the matrix elements indicate that the fuzzy cylinder has two open
ends, the zero modes surface has closed caps giving it more the shape
of a capsule. Again, it is useful to visualize the edges of the graph
with a closed string that is emerging at the lowest point $0$, which
is a border of the graph, propagating along the inner points and vanishing
at the other border point. It should be noted that the zero modes
surface has a radius smaller than $1$, although due to $s=1$ one
would expect a radius of $1$.

It is also interesting to investigate the zero modes surface of a
fuzzy cylinder with one edge reversed. In this case, numerical studies
with the computer program of the appendix show that the zero modes
surface disconnects into two pieces. For example, the zero modes surface
corresponding to a graph with three nodes and two edges pointing away
from each other or towards each other with $x_{i}=y_{i}=0$, $z_{i}=i$
and $s=1$ looks like an 2p orbital along the $z$-axis.

\subsection{Deformed fuzzy cylinder}

We now can modify the fuzzy cylinder by moving the nodes (or the corresponding
points) in space and by altering the string radii. In Fig. 3, a graph
and its zero modes surface for a cylinder of length $N=6$ is shown,
where we have set the radii of the outermost edges to $2$ and where
we have moved the two inner node by one unit in $x$-direction. In
the graph, we have arranged the nodes according to their $x$ and
$z$-coordinates, which are indicated below and to the left of the
graph.

\begin{figure}

\begin{tabular}{>{\centering}m{7cm}>{\centering}m{7cm}}
\begin{tikzpicture}
	\tikzstyle{N}=[draw,circle,fill=white,minimum size=4pt,inner sep=0pt]

	\node at (0pt,0pt)  {$0$};
	\node at (0pt,20pt) {$1$};
	\node at (0pt,40pt) {$2$};
	\node at (0pt,60pt) {$3$};
	\node at (0pt,90pt) {$4.5$};
	\node at (0pt,110pt) {$5.5$};

	\node[N] (n1) at (40pt,0pt) {}; 
	\node[N] (n2) at (40pt,20pt) {} edge[<-] (n1);
	\node[N] (n3) at (60pt,40pt) {} edge[<-] (n2);
	\node[N] (n4) at (60pt,60pt) {} edge[<-] (n3);
	\node[N] (n5) at (40pt,90pt) {} edge[<-] (n4);
	\node[N] (n6) at (40pt,110pt) {} edge[<-] (n5);

	\node at (40pt,-20pt) {$0$};
	\node at (60pt,-20pt) {$1$};

	\node at (80pt,10pt) {$s=2$}; 
	\node at (80pt,30pt) {$s=1$}; 
	\node at (80pt,50pt)  {$s=1$}; 
	\node at (80pt,75pt)  {$s=1$};
	\node at (80pt,100pt)  {$s=2$};

\end{tikzpicture} & \includegraphics[scale=0.5]{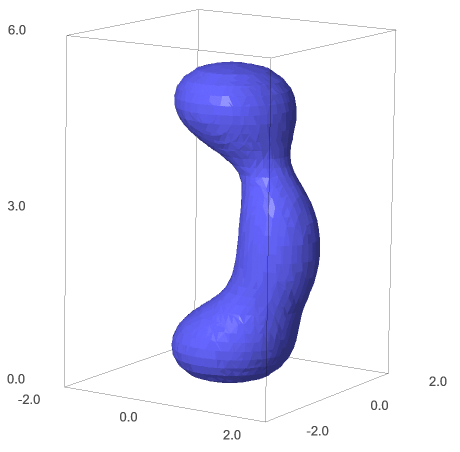}\tabularnewline
\end{tabular}\protect\caption{Deformed fuzzy cylinder}

\end{figure}
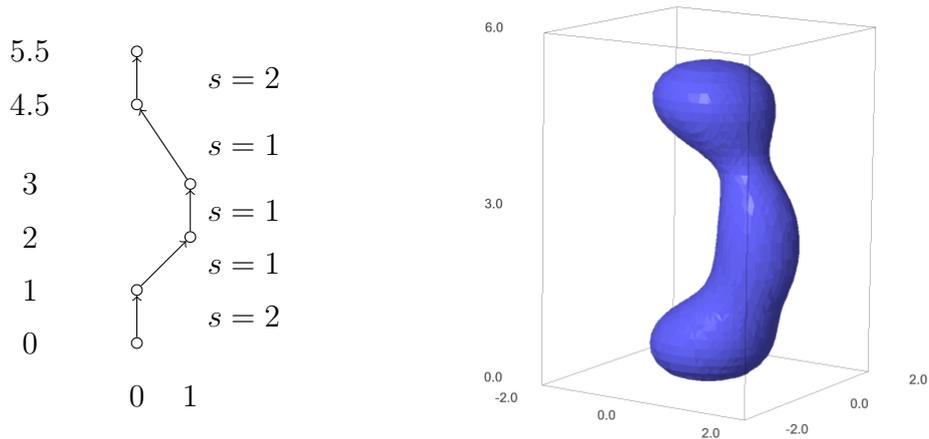

The corresponding matrices are{\small{}
\begin{equation}
V=\left(\begin{array}{cccccc}
0 & 2\\
 & 0 & 1\\
 &  & 1 & 1\\
 &  &  & 1 & 1\\
 &  &  &  & 0 & 2\\
 &  &  &  &  & 0
\end{array}\right)\ Z=\left(\begin{array}{cccccc}
0\\
 & 1\\
 &  & 2\\
 &  &  & 3\\
 &  &  &  & 4.5\\
 &  &  &  &  & 5.5
\end{array}\right)
\end{equation}
}The edge between $z=3$ and $z=4.5$ has been made longer as the
other edges. One sees that elongating the length of an edge results
in a smaller radius of the zero modes surface between the two corresponding
points. With the computer program, one also can see that there exists
an edge length, where the upper part of the zero modes surface disconnects
from the lower part.

\subsection{Fuzzy sphere}

The fuzzy sphere \cite{Steinacker:2011,Madore:1991} is defined by
the representations of $SU(2)$. With $X_{i}=L_{i}/J$ where the $L_{i}$
form the $2j+1$-dimensional representation of the $SU(2)$ and $J=\sqrt{j(j+1)}$
is the root of the Casimir, we can define $V=X_{1}+iX_{2}$ and $Z=X_{3}$.
This results in

\begin{eqnarray}
Z & = & \frac{1}{\sqrt{j(j+1)}}\sum_{m=-j}^{j}m\left|m\right\rangle \left\langle m\right|\\
V & = & \sum_{m=-j}^{j}\sqrt{1-\frac{m(m-1)}{j(j+1)}}\left|m\right\rangle \left\langle m+1\right|
\end{eqnarray}

The graph of the fuzzy sphere is a simple graph of $N=2j+1$ nodes
which are interconnected by $2j$ edges in a line. The node $m$ of
the graph is label with $(0,0,m/J)$ and the edge interconnecting
node $m$ with node $m+1$ is labeled with $s=\sqrt{1-m(m-1)/J^{2}}$.
The fuzzy sphere can be seen as a fuzzy cylinder, which string radii
are adjusted, such they fill out a sphere surface. Fig. 4 shows the
graph of the fuzzy sphere with $j=2$, where we have visualized the
parameters $s$ with circles of radius $s$. We have omitted a picture
of the zero modes surface, which is a sphere.

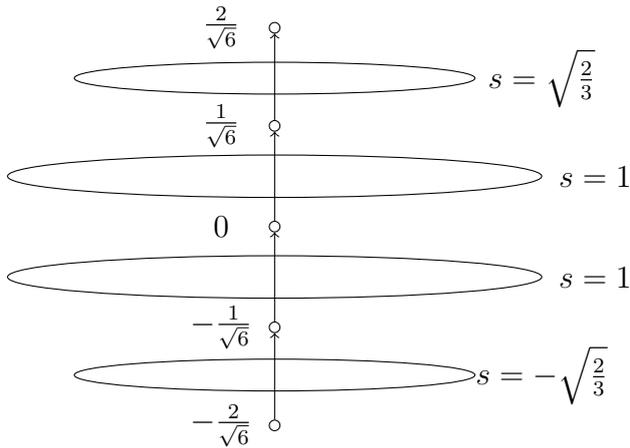
\begin{figure}
\begin{tikzpicture}
	\tikzstyle{N}=[draw,circle,fill=white,minimum size=4pt,inner sep=0pt]

	\node at (40pt,-75pt)  {$-\frac{2}{\sqrt{6}}$}; 
	\node at (40pt,-38pt) {$-\frac{1}{\sqrt{6}}$}; 
	\node at (40pt,0pt) {$0$};  
	\node at (40pt,38pt) {$\frac{1}{\sqrt{6}}$};  
	\node at (40pt,75pt) {$\frac{2}{\sqrt{6}}$};  

	\node[N] (n1) at (60pt,-75pt) {}; 
	\node[N] (n2) at (60pt,-38pt) {} edge[<-] (n1);
	\node[N] (n3) at (60pt,0pt)  {} edge[<-] (n2);
	\node[N] (n4) at (60pt,38pt) {} edge[<-] (n3);
	\node[N] (n5) at (60pt,75pt) {} edge[<-] (n4);

	\node at (160pt,-56pt) {$s=-\sqrt{\frac{2}{3}}$}; 
	\node at (180pt,-19pt) {$s=1$}; 
	\node at (180pt,19pt) {$s=1$}; 
	\node at (160pt,56pt) {$s=\sqrt{\frac{2}{3}}$}; 

	\draw (60pt,-56pt) ellipse [x radius=75pt, y radius=6pt];
	\draw (60pt,-19pt) ellipse [x radius=100pt, y radius=8pt];
	\draw (60pt,19pt) ellipse [x radius=100pt, y radius=8pt];
	\draw (60pt,56pt) ellipse [x radius=75pt, y radius=6pt];

\end{tikzpicture}

\protect\caption{Fuzzy sphere}
\end{figure}

\subsection{Fuzzy torus}

We continue with fuzzy spaces of genus 1. The usually described fuzzy
torus, which is defined by clock and shift matrices (see, for example
\cite{Steinacker:2011}), is based on a surface embedded in the four
sphere. Although it is possible to define a graph with points in $\mathbb{R}^{4}$
that reproduces this torus, we rather continue with the ``deformed
fuzzy torus'' as described in \cite{Arnlind:2009}, which is naturally
embedded in $\mathbb{R}^{3}$. This fuzzy torus is represented by
the matrices
\begin{eqnarray}
Z & = & \sum_{n=0}^{N-1}\mu\sin\left(\frac{2\pi}{N}(n+\delta)\right)\left|n\right\rangle \left\langle n\right|=\sum_{n=0}^{N-1}z_{n}\left|n\right\rangle \left\langle n\right|\label{eq:def_torus_Z}\\
V & = & \sum_{n=0}^{N-1}\sqrt{\frac{\mu}{\text{cos}\frac{\pi}{N}}\cos\left(\frac{2\pi}{N}(n+\delta+\frac{1}{2})\right)+\nu}\left|n\right\rangle \left\langle n+1\right|=\sum_{n=0}^{N-1}s_{n}\left|n\right\rangle \left\langle n+1\right|\label{eq:def_torus_V}
\end{eqnarray}
for $\nu>\mu$. Note that in the last equation, $\left|N-1\right\rangle \left\langle N\right|$
is identified with $\left|N-1\right\rangle \left\langle 0\right|$,
i.e. $V$ is ``cyclic'', which will have interesting consequences
in the following. (This is also a property of the shift matrix of
the fuzzy torus with embedding in the four sphere.) For $\nu<\mu$
and the index $n$ restricted to real values for the factors $s_{n}$
of $V$ (the value under the square root should be non-negative),
the matrices represent a ``deformed'' fuzzy sphere. However we will
concentrate on the case of the deformed torus.

From (\ref{eq:def_torus_Z}, \ref{eq:def_torus_V}) immediately follows
that the nodes of the graph only have different $z$-coordinates,
but all lie on the $z$-axis. Due to the cyclic property of $V$,
the associated graph is cyclic and the string radii $s_{n}$ are associated
to the edges between node $n$ and node $n+1$ (where $N$ is identified
with $0$). Since the deformed fuzzy torus is based on the relation
\begin{equation}
(\frac{VV^{\dagger}+V^{\dagger}V}{2}-\nu)^{2}+Z^{2}=\mu
\end{equation}
the points $(z_{n},s_{n})$ lie substantially on a deformed circle
around the point $(0,\nu)$. In the case of tori, it is a good visualization
to draw the nodes of the graphs at the coordinates $(z_{n},s_{n})$
and to assume that the graph is a cross-section of the tori along
the half-plane defined by the $z$-axis and the positive half of the
$x$-axis. However, it is important to keep in mind that the $x$-coordinate
of the visualized node is the string radius of the following edge
and not the $x$-coordinate of the point associated to the node.

Fig. 5 shows an example with $N=6$, $\mu=1$ and $\nu=1.5$. The
$\mathbb{Z}_{6}$-symmetry of the fuzzy torus is visible.

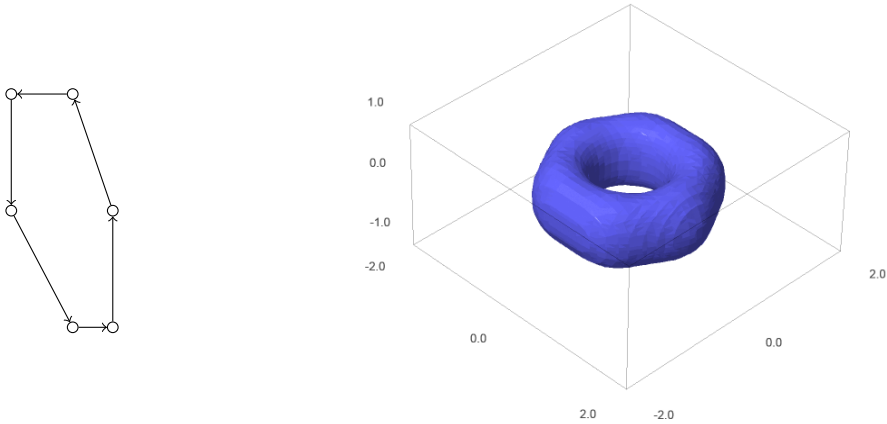
\begin{figure}
\begin{tabular}{>{\centering}m{7cm}>{\centering}m{7cm}}
\begin{tikzpicture}
	\tikzstyle{N}=[draw,circle,fill=white,minimum size=4pt,inner sep=0pt]

	\node[N] (n1) at (75pt, 0pt) {}; 
	\node[N] (n2) at (60pt, 44pt) {} edge[<-] (n1);
	\node[N] (n3) at (37pt, 44pt) {} edge[<-] (n2);
	\node[N] (n4) at (37pt,0pt) {} edge[<-] (n3);
	\node[N] (n5) at (60pt,-44pt) {} edge[<-] (n4);
	\node[N] (n6) at (75pt,-44pt) {} edge[<-] (n5) edge[->] (n1);


\end{tikzpicture} & \includegraphics[scale=0.45]{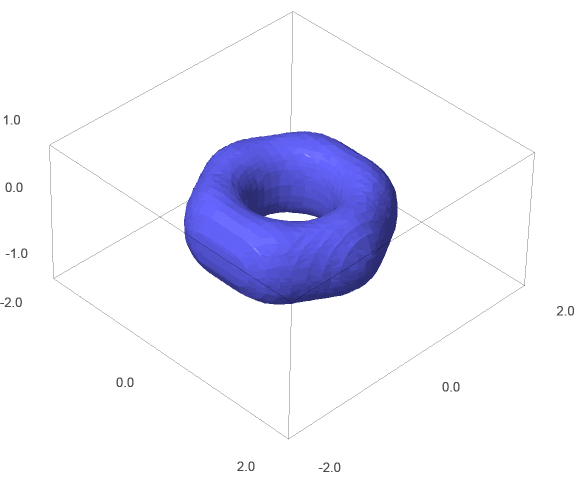}\tabularnewline
\end{tabular}\protect\caption{Deformed torus}
\end{figure}

We also can define a fuzzy version of the standard torus embedded
in $\mathbb{R}^{3}$ with the same cyclic graph as the deformed torus
but with $s_{n}=\mu\cos\left(\frac{2\pi}{N}(n+\delta)\right)+\nu$.
Fig. 6 shows an example with $N=5$, $\mu=1$ and $\nu=1.5$. Our
numerical approach shows that for the parameter $\nu$ between $1$
and 1.4, the corresponding zero modes surface does not have a hole
in the middle but is bowl shaped. 

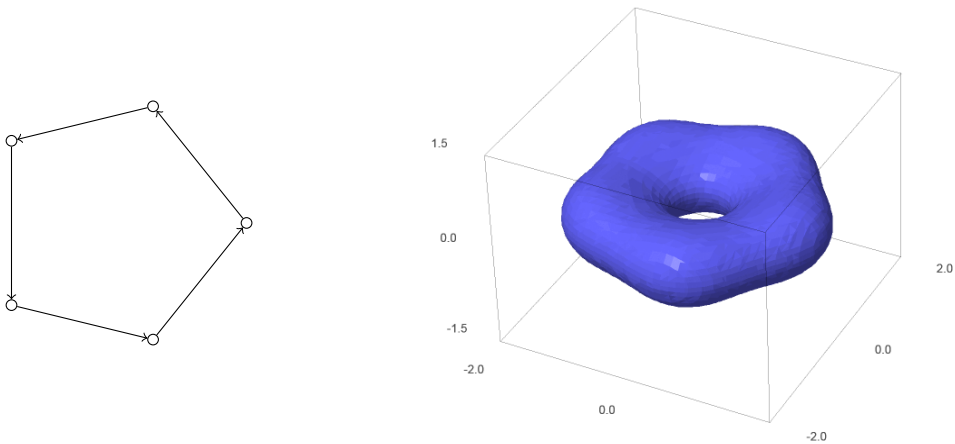
\begin{figure}
\begin{tabular}{>{\centering}m{7cm}>{\centering}m{7cm}}
\begin{tikzpicture}
	\tikzstyle{N}=[draw,circle,fill=white,minimum size=4pt,inner sep=0pt]

	\node[N] (n1) at (125pt, 0pt) {}; 
	\node[N] (n2) at (90pt, 44pt) {} edge[<-] (n1);
	\node[N] (n3) at (37pt, 31pt) {} edge[<-] (n2);
	\node[N] (n4) at (37pt, -31pt) {} edge[<-] (n3);
	\node[N] (n5) at (90pt,-44pt) {} edge[<-] (n4) edge[->] (n1);


\end{tikzpicture} & \includegraphics[scale=0.45]{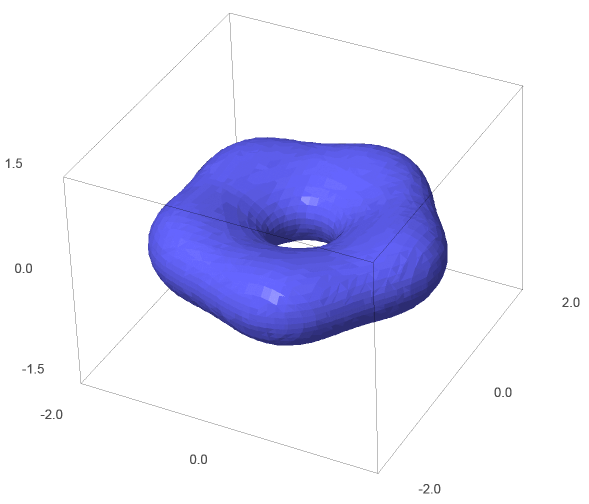}\tabularnewline
\end{tabular}\protect\caption{Standard torus}
\end{figure}

A very simple fuzzy torus can be formed from two fuzzy cylinders of
different diameter that are connected at their ends. Fig. 7 shows
a torus with $z_{n}=0,1,2,3,2,1$ and $s_{n}=1.5,1.5,1.5,4,4,4$.

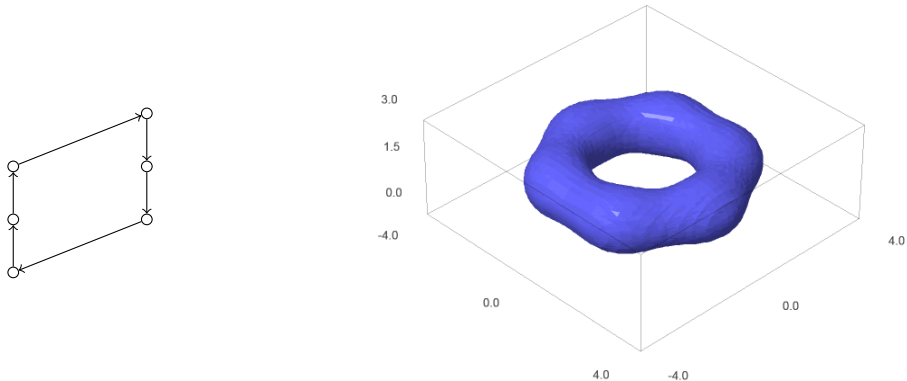
\begin{figure}
\begin{tabular}{>{\centering}m{7cm}>{\centering}m{7cm}}
\begin{tikzpicture}
	\tikzstyle{N}=[draw,circle,fill=white,minimum size=4pt,inner sep=0pt]

	\node[N] (n1) at (30pt, 0pt) {}; 
	\node[N] (n2) at (30pt, 20pt) {} edge[<-] (n1);
	\node[N] (n3) at (30pt, 40pt) {} edge[<-] (n2);
	\node[N] (n4) at (80pt, 60pt) {} edge[<-] (n3);
	\node[N] (n5) at (80pt, 40pt) {} edge[<-] (n4);
	\node[N] (n6) at (80pt, 20pt) {} edge[<-] (n5) edge[->] (n1);


\end{tikzpicture} & \includegraphics[scale=0.45]{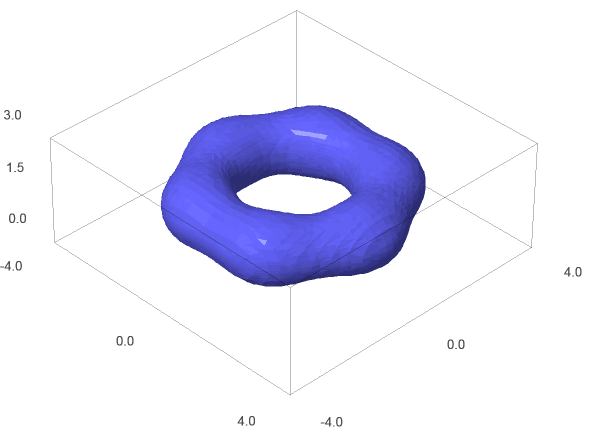}\tabularnewline
\end{tabular}\protect\caption{Simple torus}
\end{figure}

In general, with a spacing for $z_{n}$ about $1$, when the innermost
$s_{n}$ approach $1$, the zero modes surface of the tori become
more and more a bowl shaped topological sphere. When the $s_{n}$
are getting bigger, the zero modes surface decays into $N$ topological
spheres, which centers are aligned equidistant on a circle. For even
bigger $s_{n}$, the zero modes surface vanishes (at least as shown
by the computer program). The question arises, whether there is still
a manifold of minimal dispersion and how this manifold is shaped.

\subsection{From cylinder to torus}

It is interesting to start with a cylinder and to interconnect the
uppermost node with the lowest node by an additional edge. Fig. 8
shows a transformation of a fuzzy torus based on a graph with two
nodes at $z=0,1,2$, which are interconnected with two edges labeled
by $s=1$. From the left upper plot to the right the string radius
$s_{2}$ of an edge between the nodes at $z=0$ and $z=2$ is increased
from $0$ to $16$ in steps of $2$. One sees that the fuzzy cylinder
transforms in three (in general $N$) small (topological) spheres,
which reunite and form a fuzzy torus, when the string radius $s_{2}$
increases. Furthermore, before the spheres are formed, according to
the computer program, the zero modes surface vanishes between $s_{2}=1$
and $s_{2}=2$. 

The increasing of the string radius $s_{2}$ was explained in \cite{Berenstein:2012}
as adding strings with maximal angular momentum to the original fuzzy
space.

\begin{figure}

\begin{tabular}{ccc}
\includegraphics[scale=0.3]{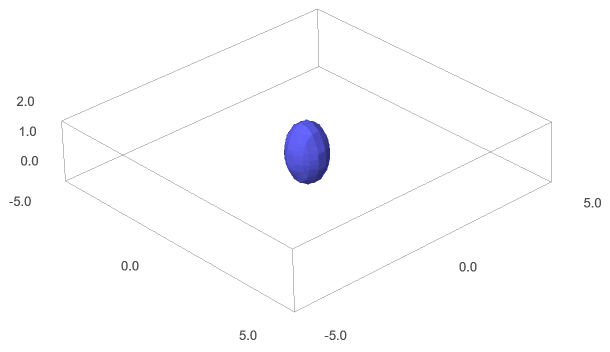} & \includegraphics[scale=0.3]{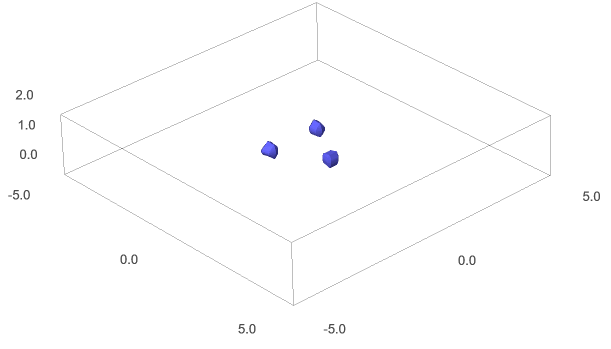} & \includegraphics[scale=0.3]{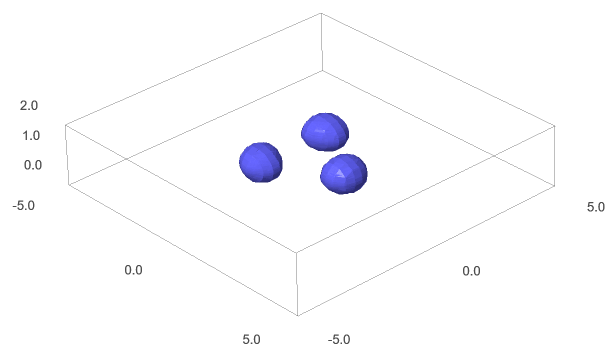}\tabularnewline
\includegraphics[scale=0.3]{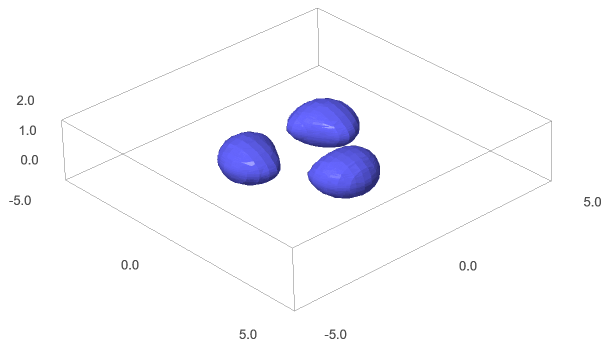} & \includegraphics[scale=0.3]{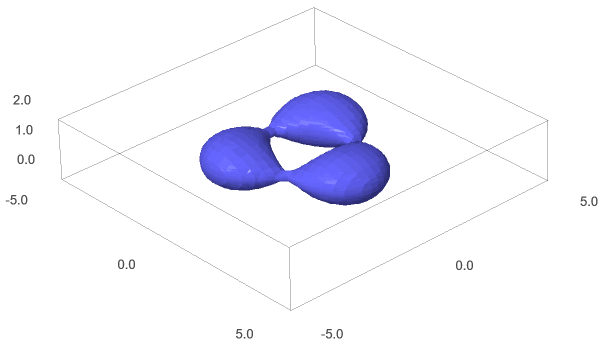} & \includegraphics[scale=0.3]{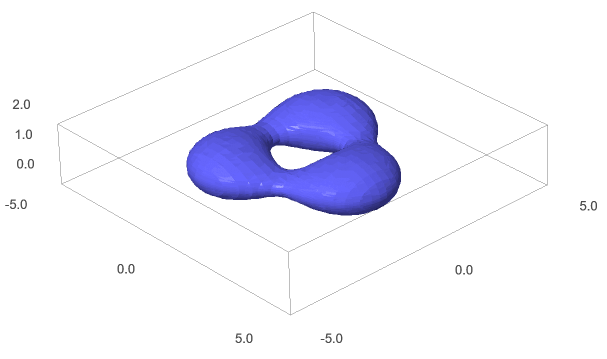}\tabularnewline
\includegraphics[scale=0.3]{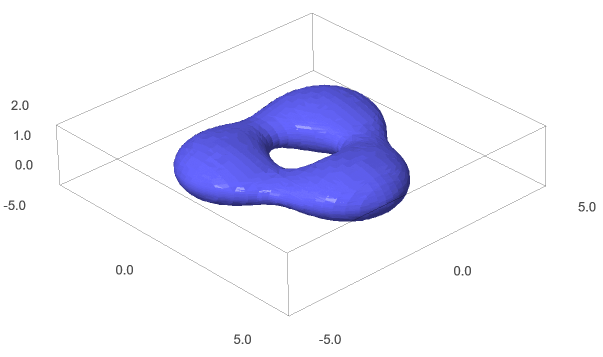} & \includegraphics[scale=0.3]{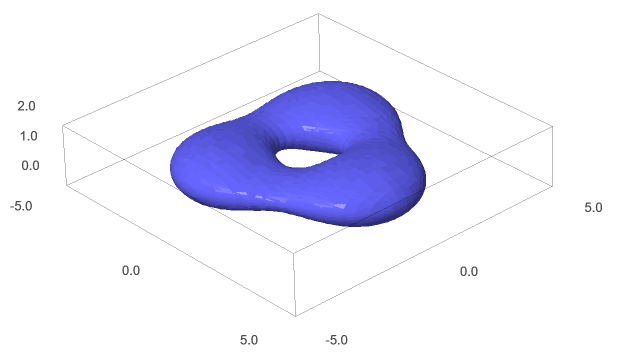} & \includegraphics[scale=0.3]{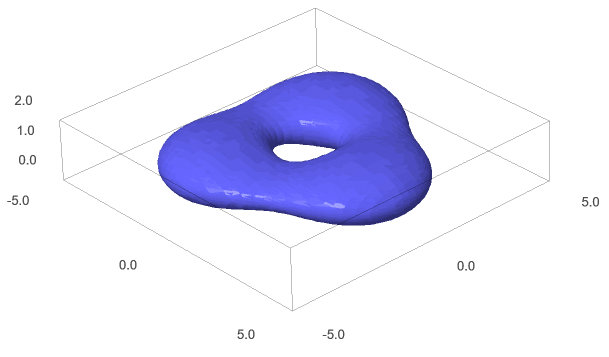}\tabularnewline
\end{tabular}\protect\caption{Cylinder transformation}

\end{figure}

\subsection{Fuzzy plane }

In three dimensions, with $X=\frac{1}{2}(V+V^{\dagger})$ and $Y=\frac{1}{2i}(V-V^{\dagger})$
the fuzzy plane can be described by 

\begin{equation}
V=\sum_{n\in\mathbb{N}_{0}}\sqrt{n+1}\left|n\right\rangle \left\langle n+1\right|,\ Z=0
\end{equation}
To be consistent, we have renamed the usual lowering and raising operators
$A$ and $A^{\dagger}$ to $V$ and $V^{\dagger}$. This infinite
dimensional matrices are related to a graph (see Fig. 9) with $\mathbb{N}_{0}$
nodes, in which the points $(x,y,z)$ of all nodes are zero. The $n+1$.th
node is connected with the $n$.th node via an edge labeled with $\sqrt{n+1}$.
Every node except the first node is connected with two other nodes. 

\begin{figure}
\begin{tikzpicture}
	\tikzstyle{N}=[draw,circle,fill=white,minimum size=4pt,inner sep=0pt]

	\node[N] (n1) at (0pt,0pt) {}; 
	\node[N] (n2) at (40pt,0pt) {} edge[<-] node[anchor=north] {$\sqrt{1}$} (n1); 
	\node[N] (n3) at (80pt,0pt) {} edge[<-] node[anchor=north] {$\sqrt{2}$} (n2); 
	\node[N] (n4) at (120pt,0pt) {} edge[<-] node[anchor=north] {$\sqrt{3}$} (n3); 
	\node[N] (n5) at (160pt,0pt) {} edge[<-] node[anchor=north] {$\sqrt{4}$} (n4);
	\node (n6) at (200pt,0pt) {} edge[<-] node[anchor=north] {$\sqrt{5}$} (n5);

	\node at (230pt,0pt) {$\cdots$};

\end{tikzpicture}

\protect\caption{Graph of the fuzzy plane}
\end{figure}
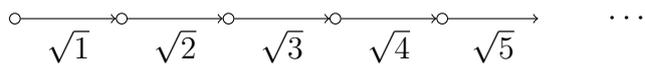

For calculating the zero modes surface, the zero eigenvalues of 
\begin{equation}
D=\left(\begin{array}{cc}
-z & V^{\dagger}-\bar{v}\\
V-v & z
\end{array}\right)
\end{equation}
have to be calculated, which has been done in \cite{Badyn:2015}.
Each eigenstate $\left|v\right\rangle $ of the lowering operator
$V$ with $V\left|v\right\rangle =v\left|v\right\rangle $ (which
defines the coherent states of the fuzzy plane) can be used to define
a zero eigenvector for $D$ for $z=0$
\begin{equation}
\left|\psi\right\rangle =\left(\begin{array}{c}
\left|v\right\rangle \\
0
\end{array}\right)
\end{equation}
Since for every $v\in\mathbb{C}$ there is a coherent state $\left|\psi\right\rangle =\exp\left(vV^{\dagger}-\bar{v}V\right)\left|0\right\rangle $,
the zero modes surface is the plane $z=0$. As mentioned above, if
one projects out a fuzzy disc from the fuzzy plane, the zero modes
manifold reduces to the origin.

\subsection{Vertices}

As we have seen above, fuzzy spaces with genus 0 are based on a graph
with two open ends, while fuzzy spaces with genus 1 are based on a
simple closed graph forming a loop. We will now extend the graph by
a branching, which results in a fuzzy space in the form of a vertex. 

Fig. 10 shows the most simple, basic vertex comprising one node, to
which three other nodes are connected. To better emphasize the vertex
structure, we have reduced $s$ for every edge to $\frac{2}{3}$. 

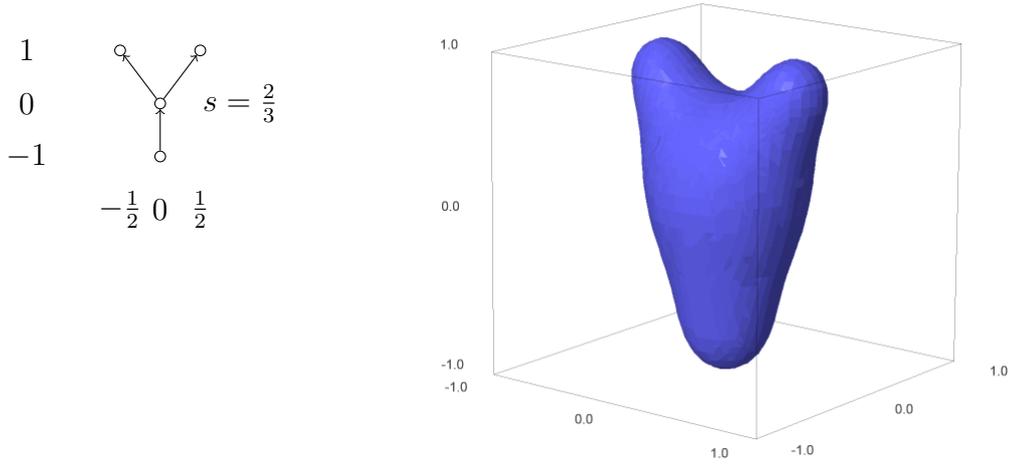
\begin{figure}
\begin{tabular}{>{\centering}p{7cm}>{\centering}m{7cm}}
\begin{tikzpicture}
	\tikzstyle{N}=[draw,circle,fill=white,minimum size=4pt,inner sep=0pt]

	\node at (0pt,0pt)  {$-1$};
	\node at (0pt,20pt) {$0$};
	\node at (0pt,40pt) {$1$};
	
	\node[N] (n1) at (50pt,0pt) {}; 
	\node[N] (n2) at (50pt,20pt) {} edge[<-] (n1);
	\node[N] (n3) at (35pt,40pt) {} edge[<-] (n2);
	\node[N] (n4) at (65pt,40pt) {} edge[<-] (n2);

	\node at (35pt,-20pt) {$-\frac{1}{2}$};
	\node at (50pt,-20pt) {$0$};
	\node at (65pt,-20pt) {$\frac{1}{2}$};

	\node at (80pt,20pt) {$s=\frac{2}{3}$};

\end{tikzpicture} & \includegraphics[scale=0.5]{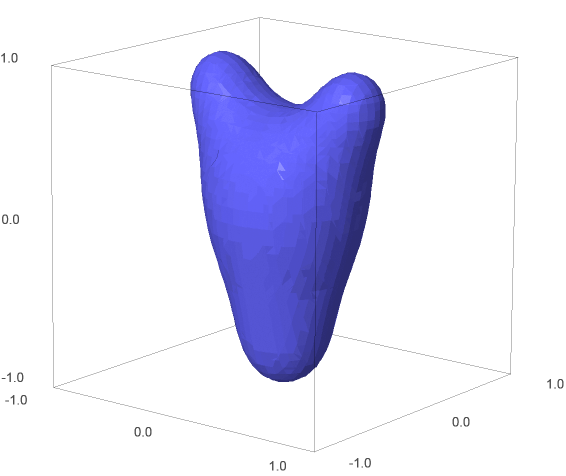}\tabularnewline
\end{tabular}

\protect\caption{Basic vertex}
\end{figure}

The corresponding matrices are

{\small{}
\begin{equation}
V=\left(\begin{array}{cccc}
0 & \frac{2}{3}\\
 & 0 & \frac{2}{3} & \frac{2}{3}\\
 &  & \frac{1}{2}\\
 &  &  & -\frac{1}{2}
\end{array}\right)\ Z=\left(\begin{array}{cccc}
-1\\
 & 0\\
 &  & 1\\
 &  &  & 1
\end{array}\right)
\end{equation}
}{\small \par}

For $s<\frac{1}{2}$, the three branches of the vertex detach from
the central part and the zero modes surface reduces to four topological
spheres. Furthermore, the direction of the edges is important. If
one of the edges is reversed, then the corresponding part of the zero
modes surface detaches from the rest of the zero modes surface.

To the free ends of the basic vertex further nodes can be attached.
The result with one further node per branch is shown in Fig. 11.

\begin{figure}
\begin{tabular}{>{\centering}p{7cm}>{\centering}m{7cm}}
\begin{tikzpicture}
	\tikzstyle{N}=[draw,circle,fill=white,minimum size=4pt,inner sep=0pt]

	\node at (0pt,0pt)  {$-2$};
	\node at (0pt,20pt)  {$-1$};
	\node at (0pt,40pt) {$0$};
	\node at (0pt,60pt) {$1$};
	\node at (0pt,80pt) {$2$};

	\node[N] (n1) at (100pt,0pt) {}; 
	\node[N] (n2) at (100pt,20pt) {} edge[<-] (n1);
	\node[N] (n3) at (100pt,40pt) {} edge[<-] (n2);
	\node[N] (n4) at (120pt,60pt) {} edge[<-] (n3);
	\node[N] (n5) at (140pt,80pt) {} edge[<-] (n4);
	\node[N] (n6) at (80pt,60pt) {} edge[<-] (n3);
	\node[N] (n7) at (60pt,80pt) {} edge[<-] (n6);

	\node at (60pt,-20pt) {$-\frac{3}{4}$};
	\node at (80pt,-20pt) {$-\frac{1}{2}$};
	\node at (100pt,-20pt) {$0$};
	\node at (120pt,-20pt) {$\frac{1}{2}$};
	\node at (140pt,-20pt) {$\frac{3}{4}$};

	\node at (150pt,40pt) {$s=\frac{2}{3}$};

\end{tikzpicture} & \includegraphics[scale=0.5]{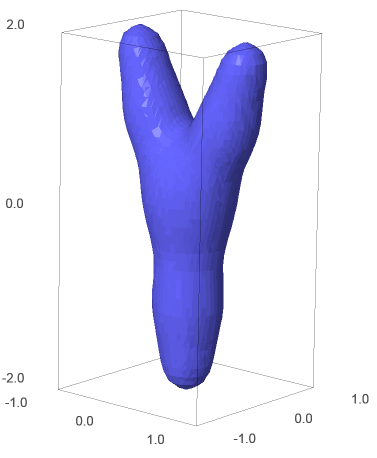}\tabularnewline
\end{tabular}

\protect\caption{Extended vertex}
\end{figure}

The corresponding matrices are

{\small{}
\begin{equation}
V=\left(\begin{array}{ccccccc}
0 & \frac{2}{3}\\
 & 0 & \frac{2}{3}\\
 &  & 0 & \frac{2}{3} &  & \frac{2}{3}\\
 &  &  & \frac{1}{2} & \frac{2}{3}\\
 &  &  &  & \frac{3}{4}\\
 &  &  &  &  & -\frac{1}{2} & \frac{2}{3}\\
 &  &  &  &  &  & -\frac{3}{4}
\end{array}\right)\ Z=\left(\begin{array}{ccccccc}
-2\\
 & -1\\
 &  & 0\\
 &  &  & 1\\
 &  &  &  & 2\\
 &  &  &  &  & 1\\
 &  &  &  &  &  & 2
\end{array}\right)
\end{equation}
}{\small \par}

It is possible to add three infinite half cylinders to the basic node,
which results in three infinite cylinders meeting in a vertex. The
corresponding infinite dimensional matrices are

{\small{}
\begin{equation}
V=\left(\begin{array}{ccccccccccc}
0 & 0 & s & s\\
s & 0 &  &  & 0\\
0 &  & -x_{0} &  &  & s\\
0 &  &  & x_{0} &  &  & s\\
 & s &  &  & 0 &  &  & 0\\
 &  & 0 &  &  & -x_{1} &  &  & s\\
 &  &  & 0 &  &  & x_{1} &  &  & s\\
 &  &  &  & s &  &  & 0 &  &  & \ddots\\
 &  &  &  &  & 0 &  &  & -x_{2}\\
 &  &  &  &  &  & 0 &  &  & x_{2}\\
 &  &  &  &  &  &  & \ddots &  &  & \ddots
\end{array}\right)
\end{equation}
\begin{equation}
Z=z_{0}\mathrm{\,diag}\left(\begin{array}{ccccccccccc}
0 & -1 & 1 & 1 & -2 & 2 & 2 & -3 & 3 & 3 & \cdots\end{array}\right)
\end{equation}
where the $x_{1}$ encode the distance from the upper nodes from the
$z$-axis and $z_{o}$ is the vertical distance between the nodes.}{\small \par}

\subsection{Vertical tori }

With a vertex it is now possible to construct tori that are aligned
vertically, i.e. have a hole that is orthogonal to the $z$-direction.
In particular, such a torus can be defined by two vertices that are
interconnected with each other with two of their branches. As the
other horizontal tori described above, the underlying graphs are also
formed of loops. However, for the vertical tori, these loops do have
nodes with points that form real loops and are not only restricted
to the $z$-axis as the points of the horizontal tori.

A first example of a torus with a graph formed of 6 nodes is shown
in Fig. 12.

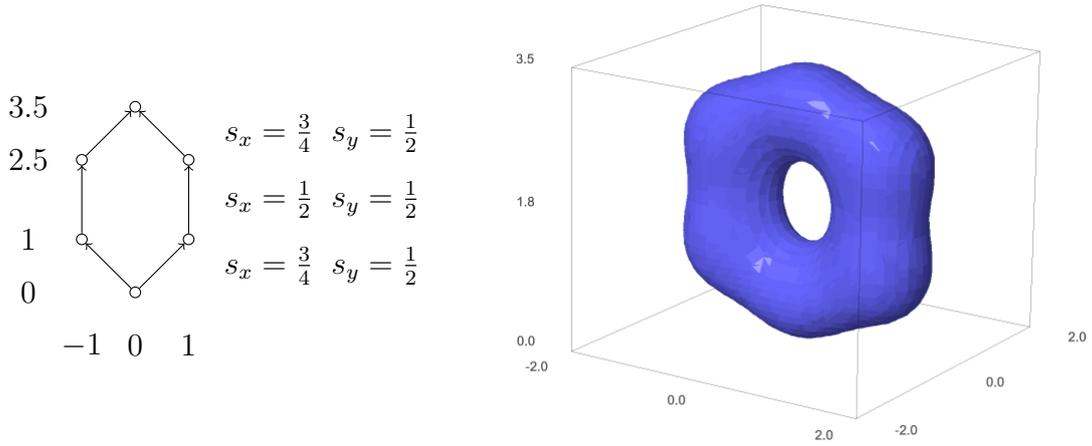
\begin{figure}
\begin{tabular}{>{\centering}m{7cm}>{\centering}m{7cm}}
\usetikzlibrary{arrows}
\begin{tikzpicture}
	\tikzstyle{N}=[draw,circle,fill=white,minimum size=4pt,inner sep=0pt]

	\node at (0pt,0pt)  {$0$};
	\node at (0pt,20pt)  {$1$};
	\node at (0pt,50pt) {$2.5$};
	\node at (0pt,70pt) {$3.5$};

	\node[N] (n1) at (40pt,0pt) {}; 
	\node[N] (n2) at (20pt,20pt) {} edge[<-] (n1);
	\node[N] (n3) at (20pt,50pt) {} edge[<-] (n2);
	\node[N] (n4) at (40pt,70pt) {} edge[<-] (n3);
	\node[N] (n5) at (60pt,50pt) {} edge[->] (n4);
	\node[N] (n6) at (60pt,20pt) {} edge[->] (n5) edge[<-] (n1);

	\node at (20pt,-20pt) {$-1$};
	\node at (40pt,-20pt) {$0$};
	\node at (60pt,-20pt) {$1$};

	\node at (90pt,10pt) {$s_x=\frac{3}{4}$};
	\node at (90pt,35pt) {$s_x=\frac{1}{2}$};
	\node at (90pt,60pt) {$s_x=\frac{3}{4}$};

	\node at (130pt,10pt) {$s_y=\frac{1}{2}$};
	\node at (130pt,35pt) {$s_y=\frac{1}{2}$};
	\node at (130pt,60pt) {$s_y=\frac{1}{2}$};

\end{tikzpicture} & \includegraphics[scale=0.5]{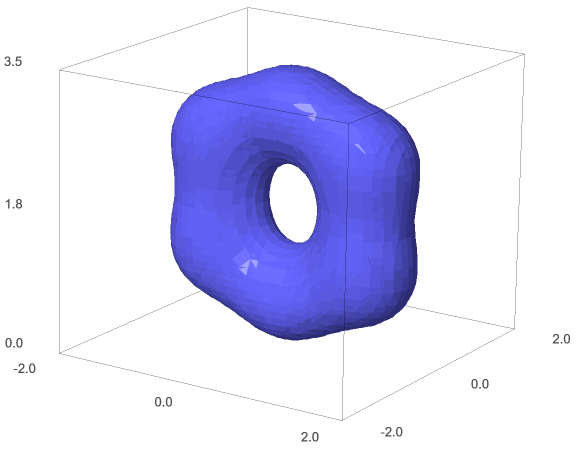}\tabularnewline
\end{tabular}

\protect\caption{Vertical torus with $\mathbb{Z}_{6}$-symmetry}
\end{figure}

{\small{}The matrices are}{\small \par}

{\small{}
\begin{equation}
X=\frac{1}{2}\left(\begin{array}{cccccc}
0 & \frac{3}{2} &  &  &  & \frac{3}{2}\\
\frac{3}{2} & -2 & 1\\
 & 1 & -2 & \frac{3}{2}\\
 &  & \frac{3}{2} & 0 & \frac{3}{2}\\
 &  &  & \frac{3}{2} & 2 & 1\\
\frac{3}{2} &  &  &  & 1 & 2
\end{array}\right)\ Y=-\frac{i}{2}\left(\begin{array}{cccccc}
0 & 1 &  &  &  & 1\\
-1 & 0 & 1\\
 & -1 & 0 & 1\\
 &  & -1 & 0 & -1\\
 &  &  & 1 & 0 & -1\\
-1 &  &  &  & 1 & 0
\end{array}\right)
\end{equation}
\begin{equation}
Z=\left(\begin{array}{cccccc}
0\\
 & 1\\
 &  & \frac{5}{2}\\
 &  &  & \frac{7}{2}\\
 &  &  &  & \frac{5}{2}\\
 &  &  &  &  & 1
\end{array}\right)
\end{equation}
}{\small \par}

As a second example, a simple fuzzy vertical torus can be formed of
only four nodes. The graph and the zero modes surface of this torus
are shown in Fig. 13. The lattice spacing for $z$ and $x$ with $5/4$
and $3/4$ has been chosen to get a zero modes surface that is more
symmetric with respect to rotations around the $y$-axis.

\begin{figure}
\begin{tabular}{>{\centering}p{7cm}>{\centering}m{7cm}}
\usetikzlibrary{arrows}
\begin{tikzpicture}
	\tikzstyle{N}=[draw,circle,fill=white,minimum size=4pt,inner sep=0pt]

	\node at (0pt,0pt)  {$-\frac{5}{4}$};
	\node at (0pt,25pt)  {$0$};
	\node at (0pt,50pt) {$-\frac{5}{4}$};

	\node[N] (n1) at (40pt,0pt) {}; 
	\node[N] (n2) at (20pt,25pt) {} edge[<-] (n1);
	\node[N] (n3) at (40pt,50pt) {} edge[<-] (n2);
	\node[N] (n4) at (60pt,25pt) {} edge[->] (n3) edge[<-] (n1);

	\node at (20pt,-20pt) {$-\frac{3}{4}$};
	\node at (40pt,-20pt) {$0$};
	\node at (60pt,-20pt) {$\frac{3}{4}$};

	\node at (90pt,50pt) {$s=1$};

\end{tikzpicture} & \includegraphics[scale=0.5]{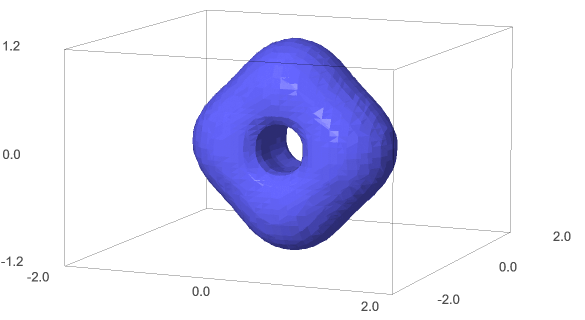}\tabularnewline
\end{tabular}

\protect\caption{Simple vertical torus with $\mathbb{Z}_{4}$-symmetry}
\end{figure}
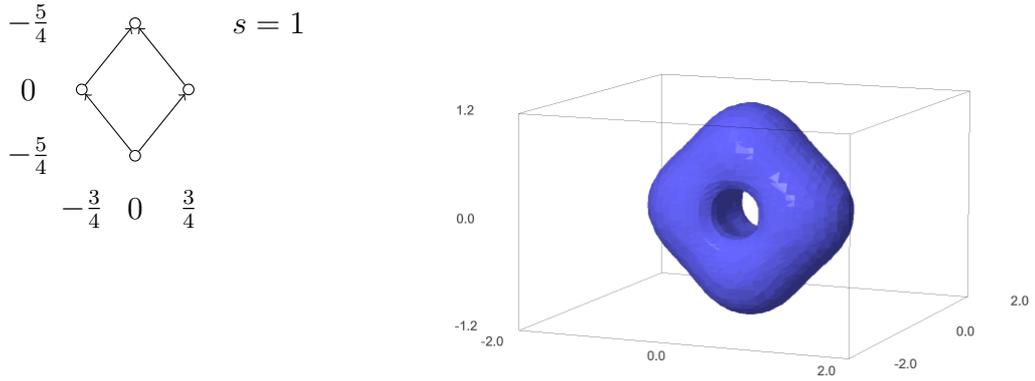

{\small{}The matrices are}{\small \par}

{\small{}
\begin{equation}
V=\left(\begin{array}{cccc}
0 & 1 & 1\\
 & \frac{3}{4} &  & 1\\
 &  & -\frac{3}{4} & 1\\
 &  &  & 0
\end{array}\right)\ Z=\frac{1}{4}\left(\begin{array}{cccc}
-5\\
 & 0\\
 &  & 0\\
 &  &  & 5
\end{array}\right)
\end{equation}
}{\small \par}

{\small{}We have sorted the nodes in such a way that the matrix $V$
is triangular. This is achieved by sorting the nodes according to
their $z$-coordinate. We will see in the following that this is useful
for extending the torus to surfaces of higher genus.}{\small \par}

\subsection{Symmetric fuzzy eight (a genus two surface)}

By interconnecting three vertices or two vertical tori we are now
able to define fuzzy spaces of higher genus (or fuzzy spaces, which
have at least zero modes surfaces of higher genus). Fig. 14 shows
the graph and the zero modes surface of a fuzzy space with matrix
dimension $N=7,$ which we call ``fuzzy eight''. The graphs comprises
two closed loops and the zero modes surface has two holes. 

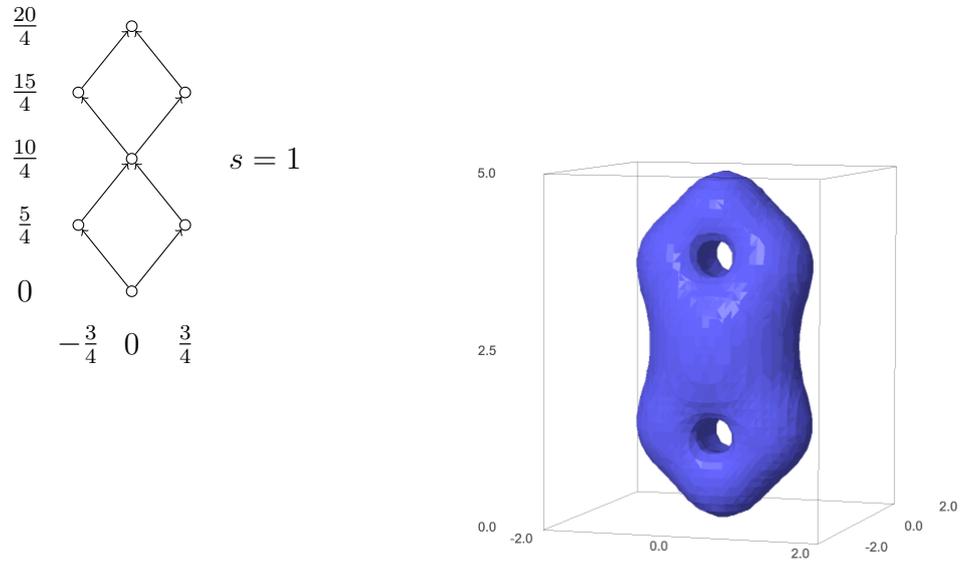
\begin{figure}
\begin{tabular}{>{\centering}p{7cm}>{\centering}m{7cm}}
\usetikzlibrary{arrows}
\begin{tikzpicture}
	\tikzstyle{N}=[draw,circle,fill=white,minimum size=4pt,inner sep=0pt]

	\node at (0pt,0pt)  {$0$};
	\node at (0pt,25pt)  {$\frac{5}{4}$};
	\node at (0pt,50pt) {$\frac{10}{4}$};
	\node at (0pt,75pt)  {$\frac{15}{4}$};
	\node at (0pt,100pt) {$\frac{20}{4}$};

	\node[N] (n0) at (40pt,0pt) {}; 
	\node[N] (n1) at (60pt,25pt) {} edge[<-] (n0);
	\node[N] (n2) at (40pt,50pt) {} edge[<-] (n1);
	\node[N] (n3) at (60pt,75pt) {} edge[<-] (n2);
	\node[N] (n4) at (40pt,100pt) {} edge[<-] (n3);
	\node[N] (n5) at (20pt,25pt) {} edge[->] (n2) edge[<-] (n0);
	\node[N] (n6) at (20pt,75pt) {} edge[->] (n4) edge[<-] (n2);

	\node at (20pt,-20pt) {$-\frac{3}{4}$};
	\node at (40pt,-20pt) {$0$};
	\node at (60pt,-20pt) {$\frac{3}{4}$};

	\node at (90pt,50pt) {$s=1$};

\end{tikzpicture} & \includegraphics[scale=0.5]{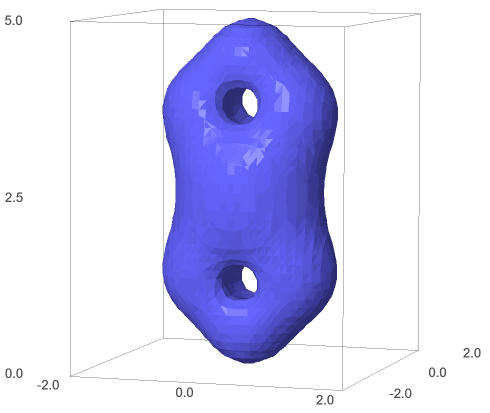}\tabularnewline
\end{tabular}

\protect\caption{Symmetric fuzzy eight}
\end{figure}

{\small{}
\begin{equation}
V=\left(\begin{array}{ccccccc}
0 & 1 & 1\\
 & \frac{3}{4} &  & 1\\
 &  & -\frac{3}{4} & 1\\
 &  &  & 0 & 1 & 1\\
 &  &  &  & \frac{3}{4} &  & 1\\
 &  &  &  &  & -\frac{3}{4} & 1\\
 &  &  &  &  &  & 0
\end{array}\right)\ Z=\frac{1}{4}\left(\begin{array}{ccccccc}
0\\
 & 5\\
 &  & 5\\
 &  &  & 10\\
 &  &  &  & 15\\
 &  &  &  &  & 15\\
 &  &  &  &  &  & 20
\end{array}\right)
\end{equation}
}{\small \par}

In Fig. 15, a section through the zero modes surface at $z=4$, i.e.
through the upper hole is shown. The intermediate (blue) line is at
$\det D_{x}=0$. The other two (red and green) lines are at $\det D_{x}$
equal to -200,0, and 200. One sees that even for $N=7$, the volume,
where the points are nearly zero, is substantially the zero modes
surface.

\begin{figure}
\includegraphics[scale=0.5]{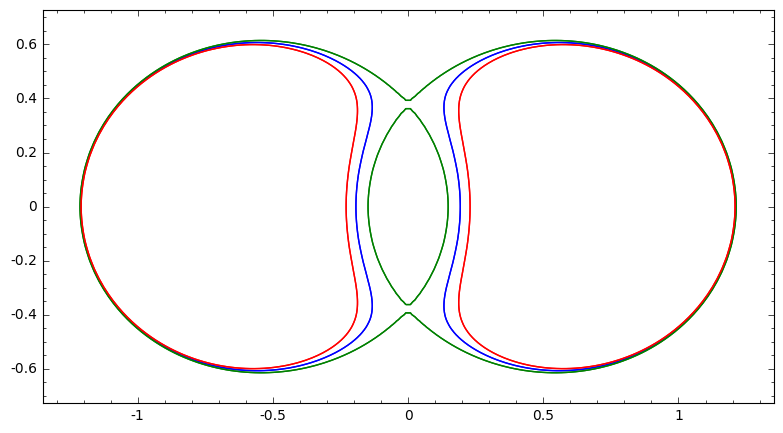}

\protect\caption{Section through fuzzy eight}
\end{figure}

Usually, when the fuzzy space is a member of a family of fuzzy spaces,
which converge in a specific limit to a classical Poisson manifold,
then the genus of the fuzzy space is defined by the genus of the classical
manifold. However up to now, there is no mathematical definition for
a genus for general non-commutative spaces. In \cite{Shimada:2003}
an interesting approach to this problem was considered. It was shown
that for the fuzzy torus, i.e. a genus 1 surface, the eigenvalues
of matrices have special properties, which there were called ``eigenvalue
sequences''. This behavior was related to Morse theory. 

In the present case of the fuzzy eight, there is per se no commutative
limit. However, due to the mapping of the graph to the diagonal $Z$-matrix,
the two branches are automatically eigenvalue sequences and one would
expect a fuzzy space of higher genus. In general, the branches of
the graph can be seen as the eigenvalue sequences of the $Z$-matrix.
On the other hand, when one numerically determines the eigenvalues
of the $X$-matrix, one sees that there are three equal eigenvalues
between one minimal and one maximal eigenvalue, i.e. there are three
branches of eigenvalues in the $X$-direction, corresponding to the
three hoses interconnecting the left and the right part of the zero
modes surface.

\subsection{Asymmetric fuzzy eight }

A fuzzy space of genus 2 with $N=6$ can be generated with the following
matrices, which graph and zero modes surface are shown in Fig. 16.

\begin{figure}
\begin{tabular}{>{\centering}p{7cm}>{\centering}m{7cm}}
\usetikzlibrary{arrows}
\begin{tikzpicture}
	\tikzstyle{N}=[draw,circle,fill=white,minimum size=4pt,inner sep=0pt]

	\node at (-20pt,0pt)  {$0$};
	\node at (-20pt,26pt) {$1.3$};
	\node at (-20pt,52pt) {$2.6$};
	\node at (-20pt,78pt) {$3.9$};

	\node[N] (n0) at (16pt, 0pt) {}; 
	\node[N] (n1) at ( 0pt,26pt) {} edge[<-] (n0);
	\node[N] (n2) at (32pt,26pt) {} edge[<-] (n0);
	\node[N] (n3) at (16pt,52pt) {} edge[<-] (n1) edge[<-] (n2);
	\node[N] (n4) at (48pt,52pt) {} edge[<-] (n2);
	\node[N] (n5) at (32pt,78pt) {} edge[<-] (n3) edge[<-] (n4);

	\node at ( 0pt,-20pt) {$0$};
	\node at (16pt,-20pt) {$0.8$};
	\node at (32pt,-20pt) {$1.6$};
	\node at (48pt,-20pt) {$2.4$};

	\node at (90pt,50pt) {$s=1$};

\end{tikzpicture} & \includegraphics[scale=0.5]{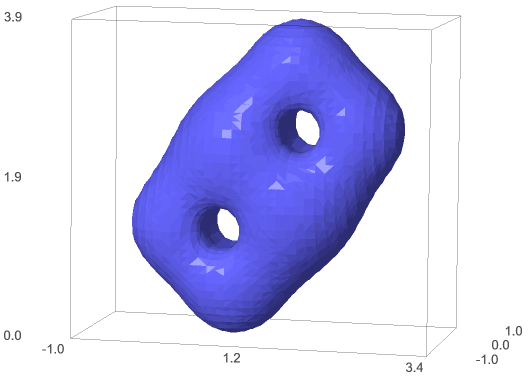}\tabularnewline
\end{tabular}

\protect\caption{Asymmetric fuzzy eight}
\end{figure}

{\small{}
\begin{equation}
V=\left(\begin{array}{cccccc}
0.8 & 1 & 1\\
 & 0 &  & 1\\
 &  & 1.6 & 1 & 1\\
 &  &  & 0.8 &  & 1\\
 &  &  &  & 2.4 & 1\\
 &  &  &  &  & 1.6
\end{array}\right)\ Z=\left(\begin{array}{cccccc}
0\\
 & 1.3\\
 &  & 1.3\\
 &  &  & 2.6\\
 &  &  &  & 2.6\\
 &  &  &  &  & 3.9
\end{array}\right)
\end{equation}
}{\small \par}

\subsection{Genus $M$ fuzzy spaces}

By extending the symmetric and asymmetric fuzzy eight, it is easy
to define matrices, which result in a two-dimensional fuzzy space
with genus $m$. An extension of the symmetric fuzzy eight results
in the following $3M+1$-dimensional matrices, which define a genus
$M$ fuzzy space.

{\small{}
\begin{equation}
V=\left(\begin{array}{ccccccccccc}
0 & 1 & 1\\
 & x_{0} &  & 1\\
 &  & -x_{0} & 1\\
 &  &  & 0 & 1 & 1\\
 &  &  &  & x_{0} &  & 1\\
 &  &  &  &  & -x_{0} & 1 & \ddots\\
 &  &  &  &  &  & 0 & \ddots\\
 &  &  &  &  &  &  & \ddots & 1 & 1\\
 &  &  &  &  &  &  &  & x_{0} &  & 1\\
 &  &  &  &  &  &  &  &  & -x_{0} & 1\\
 &  &  &  &  &  &  &  &  &  & 0
\end{array}\right)
\end{equation}
\[
Z=z_{0}\mathrm{\,diag}\left(\begin{array}{ccccccccccc}
0 & 1 & 1 & 2 & 3 & 3 & 4 & \cdots & 2M-1 & 2M-1 & 2M-2\end{array}\right)
\]
The parameters $x_{0}$ and $z_{0}$ should be about $1$ to ensure
that the holes do not close and that the components of the zero modes
surface do not disconnect.}{\small \par}

{\small{}The asymmetric fuzzy eight has an extension with $2M+1$-dimensional
matrices, which also define a genus $M$ fuzzy space.}{\small \par}

{\small{}
\begin{equation}
V=\left(\begin{array}{ccccccccc}
x_{0} & 1 & 1\\
 & 0 &  & 1\\
 &  & 2x_{0} & 1 & 1\\
 &  &  & x_{0} &  & 1\\
 &  &  &  & 3x_{0} & 1 & \ddots\\
 &  &  &  &  & 2x_{0} & \ddots & 1\\
 &  &  &  &  &  & \ddots &  & 1\\
 &  &  &  &  &  &  & (M+1)x_{0} & 1\\
 &  &  &  &  &  &  &  & Mx_{0}
\end{array}\right)
\end{equation}
\[
Z=z_{0}\mathrm{\,diag}\left(\begin{array}{ccccccccc}
0 & 1 & 1 & 2 & 2 & 3 & \cdots & M & M+1\end{array}\right)
\]
}{\small \par}

More general constructions are also possible as shown in Fig. 17.

\begin{figure}
\includegraphics[scale=0.5]{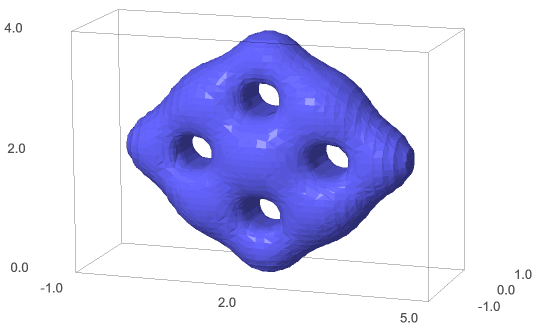}\protect\caption{Genus 4 fuzzy surface}
\end{figure}

\section{Conclusions and outlook}

We have shown, that directed graphs can be mapped to two-dimensional
fuzzy spaces that have similar properties from a topological point
of view. Simple graphs can be mapped to fuzzy spaces with zero modes
surfaces of genus 0. Graphs forming a loop can be mapped to fuzzy
spaces with zero modes surfaces having genus 1. Graphs having branchings
and forming more than one loop are mapped to fuzzy spaces with zero
modes surfaces of higher genus. Vice versa, it is possible to map
every fuzzy space defined by three Hermitian matrices to a graph.
We have also shown that this mapping is compatible with the natural
symmetry of conjugating unitary matrices with the matrices defining
the fuzzy space.

Having now the possibility to define fuzzy spaces of higher genus,
thiss can be a basis for defining sequences of graphs that result
in fuzzy spaces converging to a classical surface of higher genus
in an appropriate limit. 

A second point, which can be interesting for further investigation,
is a more covariant formulation of the mapping between graphs and
matrices. In the moment, the mapping is achieved by firstly diagonalizing
$Z$ and secondly constructing the graph from the entries of the transformed
matrices $X$ and $Y$. However, the labeling of the nodes or the
existence and labeling of an edge of the graph can be derived directly
from three general Hermitian matrices. 

We also have shown, that the symmetry of conjugating unitary matrices
gives rise to $U(1)$ lattice gauge transformations defined on the
nodes and acting on the edges. It will be interesting how lattice
gauge theories, which simply can be defined on graphs, carry over
to gauge theories on the corresponding fuzzy spaces. In this context
also the relationship with emerging gravity as described in \cite{Steinacker:2010}
can be clarified, where the $U(1)$ gauge transformations are interpreted
as coordinate transformations in the large $N$ limit.

Finally, aspects of the Standard model and the Higgs mechanism can
be formulated in terms of non-commutative geometry (see, for example,
\cite{Cammarata:1995}). As we now have a mapping from specific non-commutative
geometries to graphs, it can be possible to lift the corresponding
structures of the Standard model to graphs or generalizations thereof.

\paragraph*{Acknowledgments}

The author would like to thank Harold Steinacker for illuminating
discussions and for reading the manuscript.

\appendix

\section{SageMath program\label{sec:SageMath-program}}

The following short code listing for SageMath \cite{SageMath} shows
how to generate the graphics of the example described in section \ref{sub:The-basic-building}.
A generalization to bigger matrices or a corresponding program for
Mathematica is straightforward.

\texttt{}
\begin{lstlisting}
Z = Matrix(SR, 2, 2)
Z[0,0]= 0
Z[1,1]= 1
X = Matrix(SR, 2, 2)
X[0,1]= 1/2
X[1,0]= 1/2
Y = Matrix(SR, 2, 2)
Y[0,1]= -I*1/2
Y[1,0]= I*1/2
var('x,y,z')
X = X-x
Y = Y-y
Z = Z-z
H=block_matrix(SR,[[Z,X-I*Y],[X+I*Y,-Z]]) 
p = det(H)
g=implicit_plot3d(p, (x, -1, 1), (y, -1,1), (z, 1,1) )
g.show() 
\end{lstlisting}


\begin{thebibliography}{10}
\bibitem{Ishibashi:1997} N. Ishibashi, H. Kawai, Y. Kitazawa, and
A. Tsuchiya, \textit{A Large N reduced model as super- string}, Nucl.
Phys. B498 , 467 (1997), {[}hep-th/9612115{]}

\bibitem{Banks:1997} T. Banks, W. Fischler, S. H. Shenker, and L.
Susskind, \textit{M theory as a matrix model: A Conjecture}, Phys.
Rev. D55 , 5112 (1997), {[}hep-th/9610043{]}

\bibitem{Perelomov:1972}A. M. Perelomov, \textit{Coherent states
for arbitrary Lie group}, Comm. Math. Phys. Volume 26, Number 3 (1972),
222-236

\bibitem{Berenstein:2012}D. Berenstein, E. Dzienkowski, \textit{Matrix
embeddings on flat $R^{3}$ and the geometry of membranes}, Phys.Rev.
D86 (2012) 086001{[}arXiv:1204.2788 {[}hep-th{]}{]}

\bibitem{Ishiki:2015}G. Ishiki, \textit{Matrix Geometry and Coherent
States}, Phys.Rev. D92 (2015) no.4, 046009 {[}arXiv:1503.01230 {[}hep-th{]}{]}

\bibitem{Badyn:2015} M. H. de Badyn, J. L. Karczmarek, P. Sabella-Garnier,
K. H.-C. Yeh, \textit{Emergent geometry of membranes}, JHEP 1511 (2015)
089 {[}arXiv:1506.02035 {[}hep-th{]}{]}

\bibitem{Schneiderbauer:2016}L. Schneiderbauer, H. Steinacker, \textit{Measuring
finite Quantum Geometries via Quasi-Coherent States}, J.Phys. A49
(2016) no.28, 285301 {[}arXiv:1601.08007 {[}hep-th{]}{]}

\bibitem{Arnlind:2009}J. Arnlind, M. Bordemann, L. Hofer, J. Hoppe
and H. Shimada, \textit{Fuzzy Riemann surfaces}, JHEP \textbf{0906}
(2009) 047 {[}hep-th/0602290{]}

\bibitem{Steinacker:2011}H. Steinacker, \textit{Non-commutative geometry
and matrix models}, PoS QGQGS \textbf{2011} (2011) 004 {[}arXiv:1109.5521
{[}hep-th{]}{]}

\bibitem{Madore:1991}J. Madore, \textit{The Fuzzy sphere}, 1991,
Class.Quant.Grav. 9 (1992) 69-88

\bibitem{Shimada:2003}H. Shimada, \textit{Membrane topology and matrix
regularization}, Nucl. Phys. B \textbf{685} (2004) 297 {[}hep-th/0307058{]}

\bibitem{Lizzi:2003}F. Lizzi, P. Vitale and A. Zampini, \textit{The
Fuzzy Disc}, JHEP \textbf{0308}, 057 (2003) {[}hep-th/0306247{]}

\bibitem{Steinacker:2010}1.H. Steinacker, \textit{Emergent Geometry
and Gravity from Matrix Models: an Introduction}, Mar 2010. 57 pp.
Class.Quant.Grav. 27 (2010) 133001 {[}arXiv:1003.4134 {[}hep-th{]}

\bibitem{Cammarata:1995}G. Cammarata, Robert Coquereaux, \textit{Comments
about Higgs fields, noncommutative geometry and the standard model},
Mar 1995. 20 pp. Lect.Notes Phys. 469 (1996) 27-50 {[}hep-th/9505192{]}

\bibitem{SageMath}\textit{SageMath, the Sage Mathematics Software
System (Version 7.0)}, 2016, {[}http://www.sagemath.org{]}\end{thebibliography}
\end{document}